\documentclass[mathpazo]{aamm}
\renewcommand\thisnumber{2}
\renewcommand\thisyear {2023}
\renewcommand\thismonth{April}
\renewcommand\thisvolume{15} 

\usepackage{subfigure}
\usepackage[colorlinks=True,]{hyperref}

\begin{document}


\markboth{Shijun Liao}{Avoiding small  denominator  problems by means of the HAM}
\title{Avoiding  small  denominator  problems \\ by  means of  the homotopy analysis method}

\author[Shijun Liao]{Shijun Liao\corrauth}
\address{Center of Marine Numerical Experiment, State Key Laboratory of Ocean Engineering, Shanghai 200240, China\\
School of Naval Architecture, Ocean and Civil Engineering, Shanghai Jiaotong University, Shanghai 200240,  China}
\email{{\tt sjliao@sjtu.edu.cn} (Shijun Liao)}

%
%
%

\begin{abstract}
 The so-called  ``small   denominator   problem''  was a fundamental problem of dynamics, as pointed out by Poincar\'{e}.  Small denominators appear most commonly in perturbative theory.  The  Duffing equation is the simplest example of a non-integrable system exhibiting all problems due to small denominators.   In this paper, using the forced Duffing equation as an example, we illustrate that the famous ``small denominator problems'' never appear if a non-perturbative approach based on the homotopy analysis method (HAM), namely ``the method of directly defining inverse mapping'' (MDDiM),  is used.  The HAM-based MDDiM  provides us great freedom to directly define the inverse operator of an undetermined linear operator  so that  all small denominators can be completely avoided and besides the convergent series of multiple  limit-cycles of the  forced  Duffing equation with high nonlinearity are successfully obtained.   So, from the viewpoint of the HAM, the famous ``small denominator problems''  are only artifacts of perturbation methods.   Therefore, completely  abandoning  perturbation methods but using the HAM-based MDDiM, one would be never troubled by  ``small denominators''.  The HAM-based MDDiM  has general meanings  in mathematics and thus can be used to attack many  open problems  related  to the so-called  ``small denominators''.
\end{abstract}



\maketitle

\section{Origin of  ``small   denominator   problem'' }

 Poincar\'{e} \cite{Poincare1957} pointed  out  that the so-called  ``small   denominator   problem''  was  ``the fundamental problem of dynamics''.
 The small denominator was first mentioned by Delaunay \cite{Delaunay1867} in his 900 pages book about celestial motions  using  perturbation method.   Poincar\'{e} \cite{Poincare1957} first recognized that,  when small denominator appears, the coefficients of perturbation series may grow too large too often, threatening the convergence of the series.    As pointed out by P\'{e}rez \cite{Rodrigo2011AMS}, ``small denominators are found most commonly in the perturbative theory''.   It often appears when perturbation methods are used to solve problems in classical and  celestial mechanics \cite{Arnold1963}, fluid mechanics \cite{Iooss2005ARMA,Iooss2015},  and so on \cite{Marmi1988JPA, Chierchia1996CMP}.

What is the origin of the so-called ``small denominator problem''?   As pointed out by Giorgilli \cite{Giorgilli1998}, the Duffing equation \cite{Alex2006ND} ``is perhaps the simplest example of a non-integrable system exhibiting {\bf all} problems due to the small denominators''.
So, without loss of generality,  let us focus on  the forced Duffing equation
\begin{equation}
{\cal N}[u(t)] = u ''(t)  + 2 \; \xi \;  u'(t) + u(t) + \beta \; u^3(t) - \alpha \cos (\Omega t) = 0,   \label{geq:original}
\end{equation}
where $\cal N$ is a nonlinear operator, the prime denotes the differentiation with respect to the time $t$,  $\alpha$ and $\Omega$ is the amplitude and frequency of the external force $F=\alpha \cos(\Omega t)$,  $\xi>0$ is the resistance coefficient, and $\beta>0$  is a  physical parameter related to nonlinearity,  respectively.

As pointed out by Kartashova \cite{Kartashova2011}, ``physical classification of PDEs is based not on the form of equations, but  on the form of solutions''.    So, let us consider here the stationary periodic limit-cycle of $u(t)$ as $t\to +\infty$ of the forced Duffing equation (\ref{geq:original}),   which can be expressed in the form:
\begin{eqnarray}
u(t)
&=&   \sum_{n=1}^{+\infty}\Big\{  a_{n} \cos( \omega_{n} t ) + b_{n} \sin(\omega_{n} t) \Big\} ,\label{solution-expression}
\end{eqnarray}
where  $a_{n}$, $b_n$ are constants and
\begin{equation}
\omega_{n} = (2n-1)\Omega, \hspace{0.5cm} n \geq 1.  \label{def:omega[n]}
\end{equation}
This is mainly because the common solution \[ A \exp(-\xi t) \cos(t) + B \exp(-\xi t )\sin(t) \]  of the linear equation
\[ u''(t) + 2\xi u'(t) + u(t) = 0 \]
tends to zero as $t\to +\infty$ for arbitrary constants $A$ and $B$, and thus disappear in  the so-called ``solution-expression'' (\ref{solution-expression}) of  the limit-cycle.

Let us first show how perturbation technique \cite{Kevorkian1996,Rand1987} can bring the so-called small denominators into the above-mentioned problem.    Let $\beta$ is a small parameter and assume that $u(t)$ can be expanded in such a series
\begin{equation}
u(t) = u_0(t) + \sum_{n=1}^{+\infty} u_n(t)\; \beta^n.   \label{u:perturbation}
\end{equation}
Substituting it into (\ref{geq:original}) and  equating the like power of $\beta$, we have the  perturbation equations at different orders of $\beta$:
\begin{eqnarray}
\beta^0& : &  u_0''(t)  + 2\xi u_0'(t) + u_0(t) = \alpha \cos (\Omega t),  \label{eq:pert:0th}  \\
\beta^1& : &  u_1''(t)  + 2\xi u_1'(t) + u_1(t)  = -u^3_0(t),  \\
\beta^2& : &  u_2''(t)  + 2\xi u_2'(t) + u_2(t)  = -3 u^2_0(t) u_1(t), \\
\beta^3& : &  u_3''(t)  + 2\xi u_3'(t) + u_3(t)  = -3 u^2_0(t) u_2(t) - 3 u_0(t)u^2_1(t) ,  \\
& &  \hspace{3.0cm}\vdots  \nonumber
\end{eqnarray}

The above perturbation equations have the {\em unique} linear operator
\begin{equation}
 {\cal L}_0[u(t)] = u''(t)  + 2\xi u'(t) + u(t),   \label{L:pert}
\end{equation}
whose inverse operator ${\cal L}_0^{-1}$ reads
\begin{eqnarray}
{\cal L}_0^{-1}\left[ \cos(\omega t)\right] &=&  \frac{(1-\omega^2)\cos(\omega t)+2 \; \xi \; \omega  \sin(\omega t)}{(1-\omega^2)^2+4\xi^2\omega^2}, \label{Linv:pert:cos} \\
{\cal L}_0^{-1}\left[ \sin(\omega t)\right] &=&  \frac{(1-\omega^2)\sin(\omega t)-2 \; \xi \; \omega  \cos(\omega t)}{(1-\omega^2)^2+4\xi^2\omega^2}, \label{Linv:pert:sin}
\end{eqnarray}
where $\omega$ is a frequency.  Note that the  denominator
\[  (1-\omega^2)^2+4\xi^2\omega^2\]
becomes rather small when $\omega \to 1$ and $\xi \to 0$ so that small  denominators appear.   This is the origin of the so-called  ``small  denominator   problem''.
Thus, for Eq.~(\ref{geq:original}),  the   ``small  denominator   problem''  occurs when  $\xi$ is small and $\Omega = 1/(2n-1)$, i.e. $\omega_{n}=(2n-1)\Omega = 1$, for any a positive integer  $n\geq 1$.    Without loss of generality, let us focus on here the fixed values $\alpha=1$ and $\Omega=1/3$, say, $\omega_{2}= 3\Omega = 1$,  but investigate different values of $\xi$ and $\beta$.   In this case,  the so-called  ``resonance'' occurs when $\xi = 0$ and the ``near resonance'' occurs for a small $\xi$, corresponding to the  ``small  denominator  problem''.

In the frame of the perturbation approach, the {\em unique} initial guess is given according to (\ref{eq:pert:0th}) and (\ref{Linv:pert:cos}), say,
\begin{equation}
u_0(t) = \frac{(1-\Omega^2)\cos(\Omega t)+2 \; \xi \; \Omega  \sin(\Omega t)}{(1-\Omega^2)^2+4\xi^2\Omega^2}. \label{initial:pert}
\end{equation}
Note that small  denominator appears when $\Omega \to 1$ and $\xi \to 0$ for this perturbation initial approximation!  This is the reason why we choose $\Omega=1/3=\omega_{1}$ in this paper, otherwise the perturbation method  fails at the very beginning.

Let us first consider the perturbation method in case of $\Omega=1/3$ and $ \alpha=1$.  It is found that, when $\xi=1/100$, say, there exists the so-called ``small  denominator  problem'' for the terms $\cos(\omega_{2} t)$ and $\sin(\omega_{2} t)$,   where $\omega_2 = 3\Omega = 1$, the perturbation series is divergent  even for a small value $\beta = 0.012$, corresponding to a rather weak nonlinearity.  Thus, when the so-called ``small  denominator  problem'' appears, the perturbation method is indeed invalid in practice.   Besides,  it is found that, when $\xi=1$,  the perturbation series is also divergent for $\beta \geq 0.55$, indicating that the perturbation series is divergent for high nonlinearity even if the ``small  denominator problem''  does not occur!  Therefore, the perturbation approach  indeed does not work  for high nonlinearity and/or when the so-called ``small  denominator  problem'' occurs.  As mentioned by Arnol'd \cite{Arnold1963},  there often exist  two difficulties with perturbation method in many classical and celestial problems: (1) the appearance of small  denominator , and (2) the divergence of solution series.

The above-mentioned perturbation approach has the following disadvantages:
\begin{enumerate}

\item[(1)]  small/large physical parameters should exist;

\item[(2)]  there is no freedom to choose its linear operator;

\item[(3)]  there is no freedom to choose its initial approximation,

\item[(4)]  it is convergent only for weak nonlinearity.

\end{enumerate}
These limitations of perturbation methods are well-known.   They are the origin of the so-called  ``small  denominator  problem'' and the divergence of solution series.

Note that, due to some  historic  reasons,  the so-called ``small denominator problem'' has very close relationships with perturbation methods.     Indeed, ``small denominators are found most commonly in the perturbative theory'',  as pointed out by
P\'{e}rez \cite{Rodrigo2011AMS}.  Is perturbation method the only way to solve these problems?  What happens if we completely abandon perturbation methods ?

In this paper, we use the forced Duffing equation (\ref{geq:original})  to illustrate that the so-called ``small denominator  problem'' can never appear if we completely abandon perturbation techniques but use a non-perturbative technique, namely the homotopy analysis method (HAM) \cite{Liao1992PhD, Liao2003Book, Liao2012Book, Liao1995IJNLM, Liao1997IJNLM, Liao1999IJNLM, Liao2004AMC, Liao2007SAM,  Liao2009CNSNS, Liao2010CNSNS,Liao2016JFM, Liao2020SCPMA, Zhong2018JFM}, which is based on the basic concept ``homotopy'' in topology and can overcome all restrictions of perturbation methods.   A new HAM-based  approach is proposed, which provides us great freedom to directly define the inverse operator of an auxiliary linear operator  so that  all small denominators can be completely avoided.  Convergent series of multiple limit-cycles of the Duffing equation are successfully obtained, although the  directly defined inverse operators might be beyond the traditional mathematical theories.
Thus, from the viewpoint of the HAM, the famous ``small denominator problems''  are only artifacts of perturbation methods.   Therefore, completely  abandoning  perturbation methods, one would be never troubled by  small denominators, as illustrated below in this paper.

\section{Basic ideas of the HAM}

Can we avoid the famous ``small   denominator   problem'' in a systematic way?   The  answer  is  yes.   Here, we give an approach based on the  homotopy analysis method (HAM) \cite{Liao1992PhD, Liao2003Book, Liao2012Book},  which  can  completely  avoid  the ``small   denominator   problem''.

 ``Small denominators are found most commonly in the perturbative theory'', as pointed out by P\'{e}rez \cite{Rodrigo2011AMS}.  So,  in order to avoid ``small   denominator   problem'', we {\em must} abandon perturbation methods {\em completely}.   The  homotopy analysis method (HAM) was proposed by Liao in 1992 in his dissertation \cite{Liao1992PhD}.  Based on the basic concept  homotopy  in topology \cite{Hilton1953}, i.e. a continuous deformation, the HAM \cite{Liao2003Book, Liao2012Book, Liao1995IJNLM, Liao1997IJNLM, Liao1999IJNLM, Liao2004AMC,  Liao2007SAM,  Liao2009CNSNS,  Liao2010CNSNS, Liao2016JFM, Liao2020SCPMA, Zhong2018JFM} has the following advantages:
\begin{enumerate}
\item[(a)]  unlike perturbation techniques, the HAM works even if there exist no small/large physical parameters;

\item[(b)]  the HAM provides great freedom to choose an auxiliary  linear operator;

\item[(c)]  the HAM provides great freedom to choose an initial guess;

\item[(d)] different from other approximation methods, the HAM can guarantee the convergence of  solution series even for highly nonlinear problem.

\end{enumerate}
The HAM has been broadly used and its above-mentioned advantages have been verified and confirmed in thousands of articles by scientists and engineers all over the world \cite{Zhu2006QF,  Bouremel2007CNSNS,  Nassar2011,  Kimiaeifar2011CMA,  Ghotbi2011,  Duarte2015CSF,  Sardanyes2015,  VanGorder2017, Pfeffer2017,  Cullen2019JCP,  Sultana2019EPJP,  Massa2020,  Kaur2022JMAA,  Masjedi2022AMM,  Botton2022AMM}.
In this paper, we use the forced Duffing equation (\ref{geq:original}) as an example to illustrate how to completely avoid  the  ``small   denominator   problem'' by means of a HAM-based approach.

First, let us briefly describe the basic ideas of the HAM using (\ref{geq:original}) as an example.   Let
\begin{eqnarray}
{\cal S} = \sum_{n=1}^{+\infty} \Big[ A_{n} \cos (\omega_{n} t) + B_{n} \sin(\omega_{n} t) \Big]
\end{eqnarray}
denote a vector space,  where $\omega_{n}$ is defined by (\ref{def:omega[n]}) and $A_{n}, B_{n}$ are arbitrary constants.  Let  $u(t)\in {\cal S}$,
${\cal L}$ denote an auxiliary linear operator with the property ${\cal L}[0] =0$, which we have great freedom to choose,   $q\in[0,1]$ be a homotopy parameter,  $u_0(t)\in {\cal S}$ be an initial guess of $u(t)$, $c_0$ be a constant having no physical meanings, namely ``the convergence-control parameter'', whose value will be determined later, respectively.  Then, we construct a continuous deformation $\phi(t, q)\in{\cal S}$ from the initial guess $u_{0}\in {\cal S}$ to the solution $u(t)\in {\cal S}$ of the forced Duffing equation (\ref{geq:original}),  governed by the so-called zeroth-order deformation equation
\begin{equation}
(1-q){\cal L}\left[ \phi(t, q) -u_0(t)  \right] = c_0 \; q \; {\cal N}[\phi(t, q)],  \hspace{1.0cm} q\in[0,1], \label{geq:zeroth}
\end{equation}
where the nonlinear operator ${\cal N}[u]$ is defined by Eq.~(\ref{geq:original}).   When $q=0$, due to the property ${\cal L}[0]=0$ of the auxiliary linear operator, we have the solution
\begin{equation}
\phi(t, 0) = u_0(t)  \label{phi:q=0}
\end{equation}
of Eq.~(\ref{geq:zeroth}).  When $q=1$,  Eq. (\ref{geq:zeroth}) is exactly the same as the original equation (\ref{geq:original}), thus we have the solution
\begin{equation}
\phi(t, 1) = u(t),  \label{phi:q=1}
\end{equation}
where $u(t)$ is the solution (limiting cycle) of the original Duffing equation (\ref{geq:original}).  So, as $q$ increases from 0 to 1, $\phi(t,q)$  deforms  {\em continuously}    from the initial guess $u_0(t)$ to the solution $u(t)$ of the original nonlinear equation (\ref{geq:original}), since both of $u_{0}(t)\in{\cal S}$ and $u(t)\in{\cal S}$ can be expressed by the so-called ``solution expression'' (\ref{solution-expression}).   Then,  expanding $\phi(t,q)$ in a power series of $q$, we have according to (\ref{phi:q=0}) the homotopy-series
\begin{equation}
\phi(t,q) = u_0(t) + \sum_{k=1}^{+\infty}u_k(t) \; q^k.  \label{phi:series}
\end{equation}
Note that $\phi(t,q)\in{\cal S}$ is dependent upon the so-called convergence-control parameter $c_0$, which has no physical meanings.  Therefore, $u_{k}(t)\in{\cal S}$  $(k\geq 1)$ in (\ref{phi:series}) is also dependent upon $c_{0}$ so that the convergence radius of the series  (\ref{phi:series})   is determined by $c_0$.   Here, it should be emphasized that we have great {\em freedom} to choose the auxiliary linear operator $\cal L$, the initial guess $u_{0}$  and the convergence-control parameter $c_0$.   This is the key point of the HAM.   Assuming that the auxiliary linear operator $\cal L$, the initial guess $u_{0}$  and the convergence-control parameter $c_0$ are so properly chosen that the Maclaurin series (\ref{phi:series}) is convergent at $q=1$, we have due to (\ref{phi:q=1}) the homotopy-series solution
 \begin{equation}
u(t) = u_0(t) + \sum_{k=1}^{+\infty}u_k(t).  \label{u:series}
\end{equation}
So, even for given  auxiliary linear operator $\cal L$ and initial guess $u_{0}$,  the convergence-control parameter $c_{0}$ provides us an additional way to guarantee the convergence of the solution series, which overcomes the limitations of perturbation methods mentioned above, as illustrated below in this paper and other publications \cite{Zhu2006QF,  Bouremel2007CNSNS,  Nassar2011,  Kimiaeifar2011CMA,  Ghotbi2011,  Duarte2015CSF,  Sardanyes2015,  VanGorder2017, Pfeffer2017,  Cullen2019JCP,  Sultana2019EPJP,  Massa2020,  Kaur2022JMAA,  Masjedi2022AMM,  Botton2022AMM}.

Substituting the power series (\ref{phi:series}) into the zeroth-order deformation equation (\ref{geq:zeroth}) and  equating the like-power of $q$, we have the high-order deformation equation
\begin{equation}
{\cal L}\left[ u_m(t) -\chi_m \; u_{m-1}(t) \right] = c_0 \; R_{m-1}(t), \hspace{1.0cm} m\geq 1, \label{geq:mth}
\end{equation}
where
\begin{eqnarray}
R_k(t) &=& \left. \frac{1}{k!} \frac{d^k {\cal N}[\phi(t,q)]}{d q^k}\right|_{q=0}
\end{eqnarray}
in general and
\begin{eqnarray}
R_k(t) & = &
\left\{
\begin{array}{l}
u''_{0} +2\xi u'_0 + u_0+\beta u_0^3-\alpha \cos(\Omega t), \hspace{1.0cm}  \mbox{when $k=0$}, \\
  \\
u''_{k} +2\xi u'_k + u_k +\beta \sum\limits_{i=0}^{k}\sum\limits_{j=0}^{i} u_{k-i}u_{i-j} u_j, \hspace{0.75cm}  \mbox{when $k \geq 1$},
\end{array}
\right.  \label{def:R}
\end{eqnarray}
for the forced Duffing equation (\ref{geq:original}) considered here, and besides
\begin{equation}
\chi_m  = \left\{
\begin{array}{l}
0, \hspace{1.cm}  \mbox{when $m=1$}, \\
1, \hspace{1.cm}  \mbox{when $m >1$}.
\end{array}
\right.
\end{equation}
The general solution of the linear $m$th-order deformation equation (\ref{geq:mth}) reads
\begin{equation}
 u_m(t)  = \chi_m \; u_{m-1}(t)+c_0 \; {\cal L}^{-1}\Big[R_{m-1}(t)\Big] +\sum_{n=1}^{\mu} A_{m,n} \psi_{n}(t), \hspace{1.0cm} m\geq 1, \label{u[m]:general}
\end{equation}
where  $\mu$ is a positive integer, ${\cal L}^{-1}$ is the inverse operator of $\cal L$,  $A_{m,n}$ is an arbitary constant, and $\psi_{n}(t)\in {\cal S}$ is a base function  satisfying
\begin{equation}
{\cal L}\Big[ \sum_{n=1}^{\mu} A_{m,n}\psi_{n}(t) \Big] = 0, \hspace{0.5cm}  1\leq n \leq \mu.
\end{equation}
In other words, we have
\begin{equation}
\ker[{\cal L}] = \sum_{n=1}^{\mu} A_{m,n} \psi_{n}(t),
\end{equation}
say, the kernel of the auxiliary linear operator $\cal L$ is a vector space in dimension $\mu$.  Note that the linear part (\ref{L:pert}) of the original Duffing equation (\ref{geq:original}) is a second-order differential equation, whose kernel is a vector space in two dimension.   However, we have great freedom to choose  the auxiliary linear operator $\cal L$ and its kernel $\ker[{\cal L}]$,  as mentioned below, which might be a breakthrough in nonlinear differential equations.

The $M$th-order HAM approximation is given by
\begin{equation}
u^{*}  \approx  u_{0}(t)  +  \sum_{k=1}^{M} u_{k}(t).  \label{HAM:Mth-approximation}
\end{equation}
Since the HAM provides us great freedom to choose the initial guess $u_{0}$,  we can further use the above $M$th-order approximation as a new initial guess, say, $u_{0} = u^{*}$, to gain another $M$th-order approximation, and so on.  This provides us the $M$th-order iteration approach of the HAM.   Note that, for the HAM iteration approach, in order to avoid the exponential increment in growth of the terms in the solution expression (\ref{solution-expression}), we  eliminate the terms of $\cos(\omega_{n} t)$,  $\sin(\omega_{n} t)$ whose coefficients are less than a small value, such as $10^{-20}$ for the forced Duffing equation (\ref{geq:original}) considered in this paper.

 It should be emphasized once again that, different from all other approximation methods (including perturbation techniques), the homotopy analysis method (HAM) can guarantee the convergence of solution series by means of choosing a proper value of the so-called ``convergence-control parameter'' $c_0$.   This is the fundamental difference of the HAM from all other  approaches!  The optimal value of  the ``convergence-control parameter''  $c_0$  is determined by the minimum of the residual error square
 \begin{equation}
 {\cal E} = \int_0^T \Big( {\cal N}[u(t)] \Big)^2 dt \approx \frac{1}{K+1}\sum_{j=0}^{K} \Big({\cal N}[u(j \Delta t)]\Big)^2, \label{def:RMS}
 \end{equation}
 where $\Delta t = T/(K+1)$ is a time-step for numerical simulation, $T$ is the period of the limiting cycle for the considered problem,  $K>0$ is a large enough integer, $u(t)$ is an approximation of limiting cycle of the original equation (\ref{geq:original}), ${\cal N}$ is the nonlinear operator defined by (\ref{geq:original}), respectively.

What is the relationship between perturbation method and the HAM?    Generally speaking,  perturbation approach is often a special case of the HAM,  if  we choose the perturbation initial approximation  as $u_{0}$,  the original linear operator  as the auxiliary linear operator, i.e. ${\cal L}={\cal L}_0$,  and besides  $c_0 = -1$.  For example, when we choose  (\ref{L:pert}) as the auxiliary linear operator $\cal L$, (\ref{initial:pert}) as the initial approximation $u_{0}$, and besides set $c_{0}=-1$,    the $k$th-order deformation equation (\ref{geq:mth}) is exactly the same as the $k$th-order perturbation equation mentioned in \S~1.  Therefore, the perturbation approach can be indeed regarded as a special case of the homotopy analysis method!  However,  the perturbation approach corresponds to only one choice, but there exist {\em many} other much {\em better} choices in the frame of the HAM, which can avoid the ``small denominator problems'' completely, as illustrated below.

In summary,  the above-mentioned HAM has the following characteristics:
\begin{enumerate}
\item[(A)]  the homotopy-series (\ref{phi:series}) is expanded in the homotopy parameter $q\in[0,1]$ that has no physical meanings at all.  So, the HAM has nothing to do with any small/large physical parameters: it works no matter whether small/large physical parameters exist or not;

\item[(B)] the HAM provides us great freedom to choose its auxiliary linear operator;

\item[(C)] the HAM provides us great freedom to choose its initial approximation;

\item[(D)] the so-called convergence-control parameter $c_{0}$  has no physical meanings but can guarantee the convergence of the solution series even for high nonlinearity, as illustrated below and verified in many related publications \cite{Zhu2006QF,  Bouremel2007CNSNS,  Nassar2011,  Kimiaeifar2011CMA,  Ghotbi2011,  Duarte2015CSF,  Sardanyes2015,  VanGorder2017, Pfeffer2017,  Cullen2019JCP,  Sultana2019EPJP,  Massa2020,  Kaur2022JMAA,  Masjedi2022AMM,  Botton2022AMM}.
\end{enumerate}
Thus, the HAM can indeed overcome all limitations and restrictions of perturbation methods.

\section{How to avoid  ``small  denominator  problem'' }

As mentioned above, the HAM provides us great freedom to choose the auxiliary linear operator $\cal L$ and the initial guess $u_{0}$:  it is such kind of freedom that provides us possibility to avoid the so-called  ``small  denominator  problem'', as described below.

\subsection{Choice of the auxiliary linear operator}

Obviously, for the forced Duffing equation (\ref{geq:original}), the origin of the ``small  denominator  problem'' is mainly due to the original linear operator (\ref{L:pert}), which is unique from the viewpoint of perturbation theory.  So, in order to avoid ``small denominator problem'',  we must  abandon (\ref{L:pert}) thoroughly.   Different from other approximation techniques, the HAM provides us  great freedom to choose an auxiliary linear operator $\cal L$, as illustrated by Liao and Tan \cite{Liao2007SAM} and Liao and Zhao \cite{Liao2016NA}.  In most applications of the HAM, one often chooses a proper auxiliary linear operator $\cal L$ to gain the solution of the high-order deformation equation (\ref{geq:mth}).  However,  the freedom is so large that we can here directly {\bf define} its inverse operator ${\cal L}^{-1}$  and the kernel  of ${\cal L}$  in  (\ref{u[m]:general}).  In fact, based on the HAM,  Liao and Zhao \cite{Liao2016NA} proposed the so-called ``method of directly defining inverse mapping'', i.e. the MDDiM, which has been successfully applied to solve many types of nonlinear equations   \cite{ KV2018NA, KV2018AMC, KV2019CNA,  thesis2019Dewasurendra,  KV2021AMNS, KV2021CAM, KV2022CMA, Nave2018JBD, Nave2018JMM, Nave2019JMM, Nave2020BS,  Munjam2022IJAE, Munjam2023IJAE}.
According to the solution expression (\ref{solution-expression}) and the definition (\ref{def:R}) of $R_m(t)$,  the right-hand side of the high-order deformation equation (\ref{geq:mth}) contains terms $\cos(\omega_{n} t)$ and  $\sin(\omega_{n} t)$, where $\omega_{n} =  (2n-1) \Omega$.
 Thus, we directly {\bf define} here its inverse operator
 \begin{equation}
 {\cal L}^{-1}\Big[ A \cos(\omega_{n} t) + B \sin (\omega_{n} t) \Big] = \frac{A \cos(\omega_{n} t) + B \sin (\omega_{n} t)}{\lambda^{2}-\omega_{n}^{2}}, \hspace{0.5cm}  \left|\lambda^{2}-\omega_{n}^{2} \right| > \delta \label{L:inverse}
 \end{equation}
 and its kernel
  \begin{equation}
 {\cal L}\Big[ A' \cos(\omega_{n} t) + B' \sin (\omega_{n} t) \Big] = 0, \hspace{0.5cm}  \left|\lambda^{2}-\omega_{n}^{2} \right| \leq \delta, \label{L:kernel}
 \end{equation}
 for arbitrary constants $A,B, A', B'$, where we have great freedom to choose the two parameters $\lambda>0$ and  $\delta \geq 0$.   Note that both of $\cal L$ and ${\cal L}^{-1}$ are linear, say,
\begin{equation}
{\cal L}\Big[ A \cos(\omega_{n} t) + B \sin (\omega_{n} t) \Big] = A \; {\cal L}\Big[ \cos(\omega_{n} t) \Big] +B \; {\cal L}\Big[ \sin(\omega_{n} t) \Big],
\end{equation}
\begin{equation}
 {\cal L}^{-1}\Big[ A' \cos(\omega_{n} t) + B' \sin (\omega_{n} t) \Big] = A' \; {\cal L}^{-1}\Big[ \cos(\omega_{n} t) \Big] +B' \; {\cal L}^{-1}\Big[ \sin(\omega_{n} t) \Big],
\end{equation}
for arbitrary constants $A, B, A'$ and $B'$.  The above definitions are complete, according to the so-called ``solution-expression'' (\ref{solution-expression}).

 Let
\begin{equation}
 W_{\lambda, \delta} = \Big\{ \omega_{n} :   \left|\lambda^{2}-\omega_{n}^{2} \right| \leq \delta \Big\} \label{def:W:lambda-delta}
\end{equation}
denote a set containing all frequencies $\omega_{n}$ that satisfies $\left|\lambda^{2}-\omega_{n}^{2} \right| \leq \delta$ for a given pair of  $\lambda$ and $\delta$, where $\omega_{n} = (2n-1)\Omega$ is the frequency defined by  (\ref{def:omega[n]}).
  Then,  its inverse operator (\ref{L:inverse}) and the kernel (\ref{L:kernel})  of the auxiliary linear operator $\cal L$  can be rewritten  by
 \begin{equation}
 {\cal L}^{-1}\Big[ A \cos(\omega_{n} t) + B \sin (\omega_{n} t) \Big] = \frac{A \cos(\omega_{n} t) + B \sin (\omega_{n} t)}{\lambda^{2}-\omega_{n}^{2}}, \hspace{0.5cm}  \omega_{n} \notin W_{\lambda,\delta} \label{L:inverse:B}
 \end{equation}
 and
  \begin{equation}
 {\cal L}\Big[ A' \cos(\omega_{n} t) + B' \sin (\omega_{n} t) \Big] = 0, \hspace{0.5cm}  \omega_{n} \in W_{\lambda,\delta}, \label{L:kernel:B}
 \end{equation}
 respectively, where  $A, B, A'$ and $B'$  are arbitrary constants.   Assume that $W_{\lambda,\delta}$ contains $\kappa$ frequencies.
Then,  the kernel of the auxiliary linear operator ${\cal L}$ defined by (\ref{L:inverse}) and (\ref{L:kernel}) is a vector space in dimension $\mu = 2\kappa$, say,
 \begin{equation}
\ker \left[ {\cal L} \right] =  \sum_{\omega_{n} \in W_{\lambda,\delta}} \Big[ A_{n} \cos(\omega_{n} t)+B_{n} \sin(\omega_{n} t)\Big]. \label{ker[L]}
 \end{equation}
 for arbitrary constants $A_{n} $ and $B_{n} $.  Note that, in the frame of the HAM, we have great freedom to choose the dimension $\mu$ that is determined by $\lambda$ and $\delta$, as mentioned below.  This is completely different from the traditional mathematical  theory for a second-order differential equation.
 Thus,  the general solution of the $m$th-order deformation equation (\ref{geq:mth}) reads
 \begin{eqnarray}
 u_m(t)  &=& \chi_m \; u_{m-1}(t)+c_0 \; {\cal L}^{-1}\Big[R_{m-1}(t)\Big] \nonumber\\
 &+& \sum_{\omega_{n} \in W_{\lambda,\delta}} \Big[ A_{m,n} \cos(\omega_{n} t)+B_{m,n} \sin(\omega_{n} t)\Big], \label{u[m]:B}
 \end{eqnarray}
where $A_{m, n}$ and $B_{m, n}$ are unknown constants.  Note that ${\cal L}^{-1}\Big[R_{m-1}(t)\Big]$ can be directly obtained using the inverse operator {\bf defined} by (\ref{L:inverse}) or (\ref{L:inverse:B}), and the term  $u_{m-1}(t)$ is known.  The unknown constants $A_{m,n}$ and $B_{m,n}$ are determined via $R_{m}(t)$ in the way described below.

According to the definition (\ref{L:inverse}) or (\ref{L:inverse:B}) of the inverse operator ${\cal L}^{-1}$,  $R_{m}(t)$ can not contain the terms  $\cos(\omega_{n} t)$ and $ \sin (\omega_{n} t) $, where $\omega_{n} \in W_{\lambda,\delta}$, since there are {\bf no} definitions on them.   Substituting $u_{m}$ defined by (\ref{u[m]:B}) into $R_{m}(t)$  defined by  (\ref{def:R}), we have
\begin{equation}
R_{m}(t) =  \sum  \Big[ Q_{m,n}\cos (\omega_{n} t) + S_{m,n}\sin(\omega_{n} t) \Big].
\end{equation}
To avoid the appearance of  $\cos(\omega_{n} t)$ and $ \sin (\omega_{n} t) $ terms in the above expression, where $\omega_{n} \in W_{\lambda,\delta}$, we had to enforce the following coefficients to be zero:
\begin{equation}
Q_{m,n} = 0, \hspace{0.5cm}  S_{m,n} = 0, \hspace{1.0cm} \omega_{n} \in W_{\lambda,\delta},
\end{equation}
which give us $\mu = 2\kappa$ {\em linear} algebraic equations that determine  the $\mu=2\kappa$  unknown  coefficients  $A_{m,n}$ and $B_{m,n}$ of $u_{m}$ defined by (\ref{u[m]:B}).  In this way, we successively gain the solution $u_{m}(t)$ of the $m$th-order deformation equation (\ref{geq:mth}), where $m=1,2,3$ and so on, {\em without} any small denominators.

According to the definition (\ref{L:inverse})  or (\ref{L:inverse:B}) of the  linear  inverse operator ${\cal L}^{-1}$,  we can always choose a proper pair of $\lambda$ and $\delta$  so as to avoid the so-called  ``small  denominator  problem'', as illustrated below in \S~4.  Note that it is the HAM \cite{Liao2007SAM, Liao2016NA} that provides us such kind of great freedom.

\subsection{Choice of the initial guess\label{section:initial-guess}}

The perturbation method provides us the {\em unique} initial guess (\ref{initial:pert}), which unfortunately contains the small  denominator  $(1-\omega^{2})^{2}+4 \xi^{2} \omega^{2}$ when the frequency $\omega$ is close to 1 and $\xi$ is small.  It is well-known that perturbation approaches become invalid  when small   denominators appear.  In addition, in the frame of perturbation techniques, there is no freedom to choose initial guess.  So, we had to abandon the  initial guess (\ref{initial:pert})  of  the perturbation method.

 In the frame of the HAM, for the limit cycle of the Duffing equation (\ref{geq:original}),   all approximations should be in the form of  (\ref{solution-expression}), called the ``solution expression''.
So, the initial guess should agree with the ``solution expression'' (\ref{solution-expression}).  Besides, it should contain at least one or two lowest frequencies, such as $\cos(\omega_{1} t), \sin(\omega_{1} t)$, $\cos(\omega_{2} t)$ and $\sin(\omega_{2} t)$ for the forced Duffing equation  (\ref{geq:original}).  In addition, it should contain the kernel  (\ref{ker[L]}) of the auxiliary linear operator $\cal L$, too.
 For the forced Duffing equation (\ref{geq:original}),  considering the solution-expression (\ref{solution-expression}) and the kernel (\ref{ker[L]}) of the auxiliary linear operator $\cal L$, we choose the  initial guess in the form
 \begin{eqnarray}
 u_0(t)  &=&  \sum_{j=1}^{\gamma} \Big[a_{0,j} \cos (\omega_{j} t)+b_{0,j}\sin(\omega_{j} t)\Big]\nonumber\\
 &+&\sum_{\omega_{n} \in W_{\lambda,\delta}} \Big[ A_{0,n} \cos(\omega_{n} t)+B_{0,n} \sin(\omega_{n} t)\Big], \label{def:u[0]}
 \end{eqnarray}
 where $\gamma\geq 1$ is an integer,  and $a_{0,j},b_{0,j},  A_{0,n} ,  B_{0,n} $ are unknown constants.  Note that the same terms in the above expression should be combined.    All of these unknown constants in (\ref{def:u[0]}) are determined in the way mentioned below.

According to (\ref{def:R}), $R_{0}(t) ={\cal N}[u_{0}(t)]$ denotes the residual error of the forced Duffing equation~(\ref{geq:original}) for the initial guess $u_{0}$.
So, using the initial guess (\ref{def:u[0]}), we have
\begin{equation}
R_{0} =  \sum  \Big[ Q_{0,n}\cos (\omega_{n} t) + S_{0,n}\sin(\omega_{n} t) \Big].
\end{equation}
To avoid the appearance of the terms $\cos(\omega_{n} t), \sin(\omega_{n} t)$ in the above expression, where $\omega_{n} \in W_{\lambda,\delta}$ as defined by (\ref{def:W:lambda-delta}),
  we had to enforce the following coefficients to be zero:
\begin{equation}
Q_{0,n} = 0, \hspace{0.5cm} S_{0,n} = 0, \hspace{1.0cm} \omega_{n} \in W_{\lambda,\delta},   \label{geq:u[0]:2nd}
\end{equation}
which provides us $\mu=2\kappa$ algebraic equations.  Besides, if necessary, we had better enforce the disappearance of the base functions with the lowest frequencies, such as $\cos(\omega_{1} t), \sin(\omega_{1} t)$, $\cos(\omega_{2} t)$ and $\sin(\omega_{2} t)$,  in the above expression of $R_{0}(t)$.  In this way, all unknown constants in the initial guess (\ref{def:u[0]}) could be gained.  Note that it is a set of nonlinear algebraic equations with a few unknowns, which can be solved by means of some well-known symbolic computation software, such as the commends {\bf FindRoot} and {\bf NSolve} of {\em mathematica}.

Assume that, in the iteration approach of the HAM mentioned in \S~2, we have a known approximation $u^{*}(t)$.  Then, we choose  the initial guess
 \begin{equation}
 u_0(t)  = u^{*}(t)+\sum_{\omega_{n} \in W_{\lambda,\delta}} \Big[ A_{0,n} \cos(\omega_{n} t)+B_{0,n} \sin(\omega_{n} t)\Big], \label{def:u[0]:iteration}
 \end{equation}
where the unknown constants $A_{0,n}, B_{0,n} $ are determined by enforcing the disappearance of the kernel terms of ${\cal L}$ in $R_{0}(t)$ in the similar way as mentioned above.

In this way, we can avoid  the ``small  denominator  problem'' in the initial guess $u_{0}(t)$, which occurs for the perturbation initial guess (\ref{initial:pert}) when $\Omega \to 1$ and $\xi \to 0$.  More importantly,  a set of nonlinear algebraic equations often has multiple solutions, which might lead to multiple solutions of the limit-cycle for the forced Duffing equation (\ref{geq:original}), as described below.

Finally, we should emphasize that it is the HAM that provides us great freedom to choose the initial guess $u_{0}$.

\section{Some examples}

In this section,  let us use the forced Duffing equation (\ref{geq:original}) to illustrate the validity and novelty of the HAM approach mentioned in \S~2 and \S~3.  Without loss of generality, let us consider the case of $\alpha=1$, $\Omega = 1/3$ but various values of $\beta$ and $\xi$.  Note that $\beta$ is a measurement of the nonlinearity of the forced Duffing equation (\ref{geq:original}):  the larger the value of $\beta$, the higher the nonlinearity of the forced Duffing equation (\ref{geq:original}).

Since $\Omega=1/3$ is fixed, we always have  $\omega_{2} = 3\Omega = 1$.  Thus, from the viewpoint of perturbation techniques, the so-called  ``small  denominator  problem'' happens when $\xi$ is small, such as $10^{-4}\leq \xi \leq 10^{-2}$, so that the perturbation method fails,   as mentioned in \S~1.  However,  we illustrate here that  such kind of ``small  denominator problem'' {\bf never} appears in the frame of the HAM approach, as long as we properly choose a pair of $\lambda$ and $\delta$.

Since $\Omega=1/3$ and $\omega_{n}= (2n-1)\Omega$, we have
\begin{equation}
\omega_{1} = \frac{1}{3}, \; \omega_{2} = 1, \; \omega_{3} = \frac{5}{3}, \; \omega_{4} = \frac{7}{3}, \; \omega_{5} = 3, \; \omega_{6} = \frac{11}{3}, \; \cdots  \label{list:omega}
\end{equation}
in this paper.  In theory, there are an infinite number of ways to choose the values of $\lambda$ and $\delta$.  In this section, we just consider the following two cases:
\begin{enumerate}
\item[(a)] $\lambda=\sqrt{2}$ and $\delta=0$;

\item[(b)] $\lambda =\omega_{1}$ and $\delta = \Big| \omega_{1}^{2}-\omega^{2}_{\kappa} \Big|$ with $\kappa\geq 1$.

\end{enumerate}
All of them can completely avoid the so-called ``small  denominator  problem''.

\subsection{In case of $\lambda=\sqrt{2}$ and $\delta=0$\label{case:A}}

\begin{table}
\renewcommand\arraystretch{1.5}
\tabcolsep 0pt
\caption{Residual error square of the forced Duffing equation  (\ref{geq:original})  for $u(t)$ at different order of approximations  in case of $\alpha=1, \beta=1,\Omega=1/3$ and different values of $\xi$, given by the HAM approach in the case of $\lambda=\sqrt{2}$ and $\delta = 0$ described in \S~4.1.  }
\vspace*{-7pt}\label{table:RMS:A}
\begin{center}
\def\temptablewidth{1.0\textwidth}
{\rule{\temptablewidth}{1pt}}
\begin{tabular*}{\temptablewidth}{@{\extracolsep{\fill}}ccccc}
Order of  & $\xi = 0  $  & $\xi=10^{-4} $ & $\xi = 0.01$ & $\xi =0.1  $ \\
approximation &  \hspace{0.5cm} $c_0 = -9/10$\hspace{0.5cm}  & \hspace{0.5cm} $ c_0=-9/10$\hspace{0.5cm} & \hspace{0.5cm}$ c_0=-9/10$\hspace{0.5cm} & \hspace{0.5cm} $c_0=-8/10$\hspace{0.5cm} \\ \hline
0  	& 1.2E-3   & 1.2E-3   & 1.2E-3   & 4.9E-4 \\
1	& 4.1E-4	& 4.1E-4	& 4.1E-4	& 4.4E-4 \\
3	& 5.3E-6	& 5.3E-6	& 5.2E-6	& 2.5E-5 \\
5	& 2.4E-7	& 2.4E-7	& 2.5E-7	& 2.1E-6 \\
10	& 9.5E-11	& 9.5E-11	& 1.0E-10	& 4.6E-9 \\
15	& 4.0E-14	& 4.0E-14 & 4.7E-14	& 1.4E-11 \\
20	& 4.2E-17	& 4.2E-17	& 4.8E-17	& 5.1E-14 \\
25	& 5.7E-20	& 5.7E-20	& 6.7E-20	& 2.1E-16 \\
30	& 4.1E-23	& 4.1E-23 & 5.3E-23 & 9.1E-19 \\	
\end{tabular*}
{\rule{\temptablewidth}{1pt}}
\end{center}
\end{table}

  \begin{figure}[tb]
    \begin{center}
        \begin{tabular}{cc}
             \subfigure[]{\includegraphics[width=2.5in]{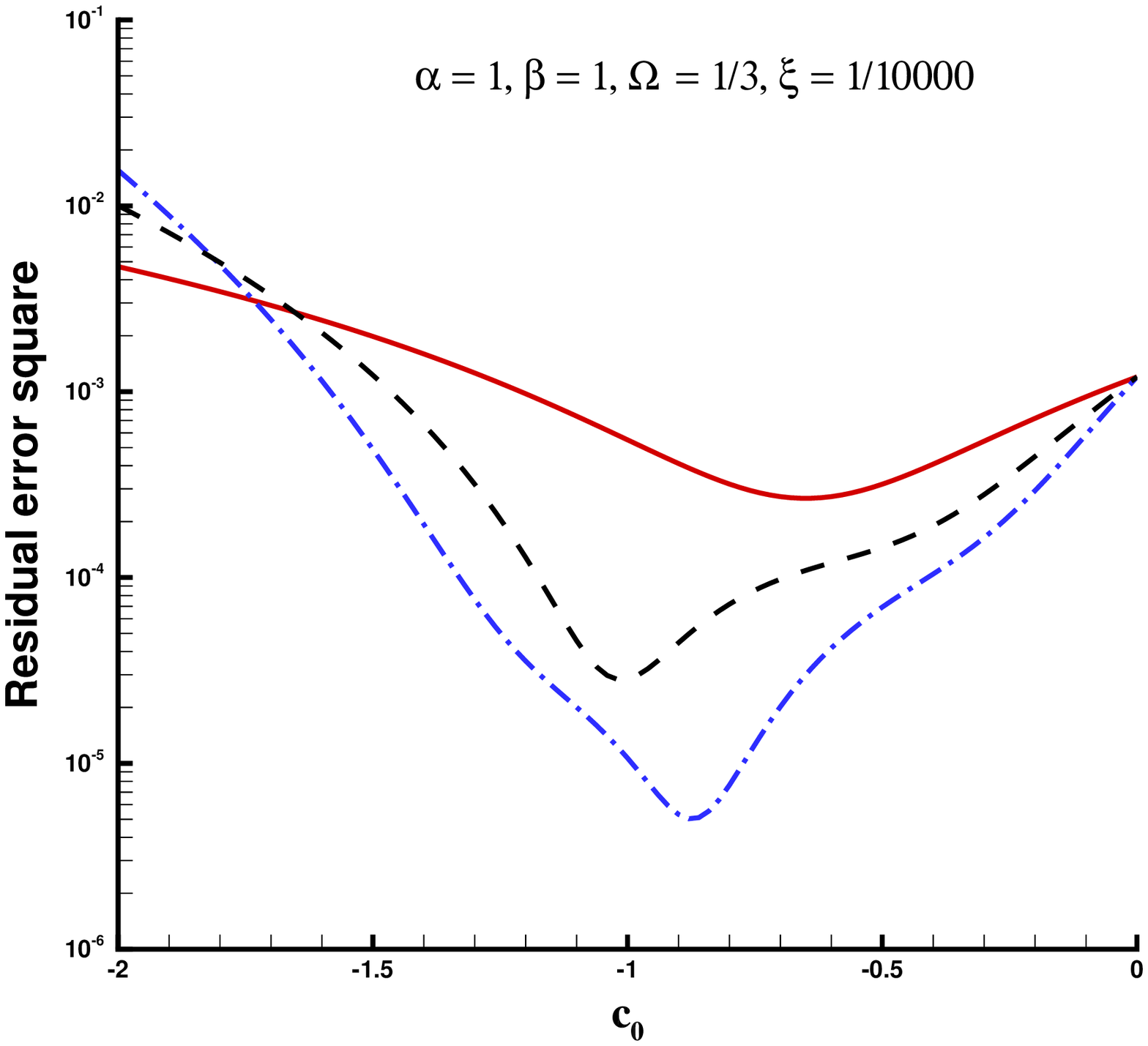}}
             \subfigure[]{\includegraphics[width=2.5in]{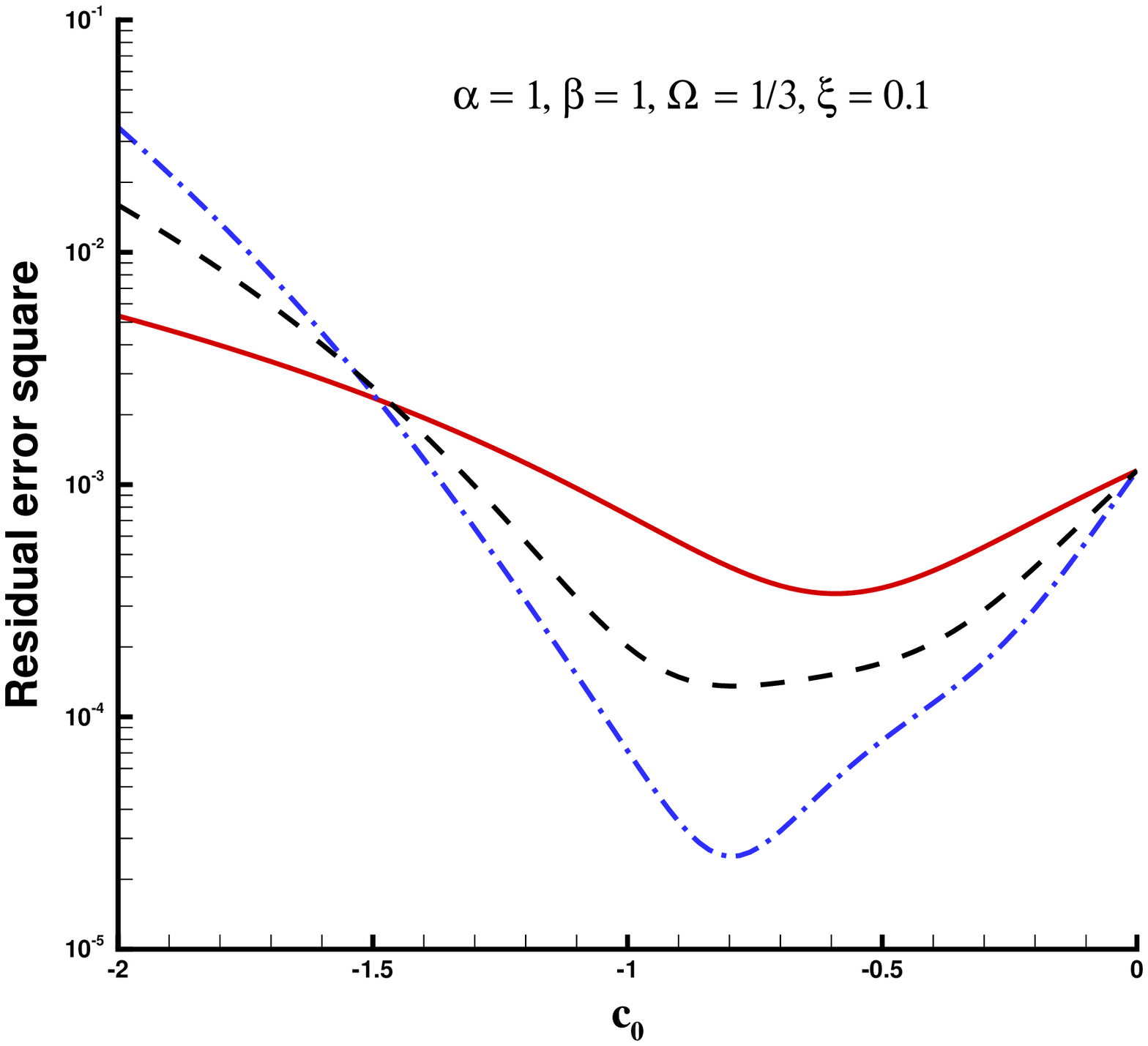}}
        \end{tabular}
 \end{center}
\caption{Residual error square of  the forced Duffing equation (\ref{geq:original}) for $u(t)$  at different orders of approximation versus the convergence-control parameter $c_0$ in case of $\alpha=1, \beta=1, \Omega=1/3$ with different values of $\xi$, given by the HAM approach in the case of $\lambda=\sqrt{2}$ and $\delta = 0$ described in \S~4.1.  Solid line: 1st-order approximation; Dashed-line: 2nd-order approximation; Dash-dotted line: 3rd-order approximation.  (a) $\xi =10^{-4}$ using  the initial guess (\ref{initial:A});  (b) $\xi=0.1$.  }
 \label{fig:RMS-A}
 \end{figure}

  \begin{figure}
    \begin{center}
        \begin{tabular}{cc}
             \subfigure[]{\includegraphics[width=2.5in]{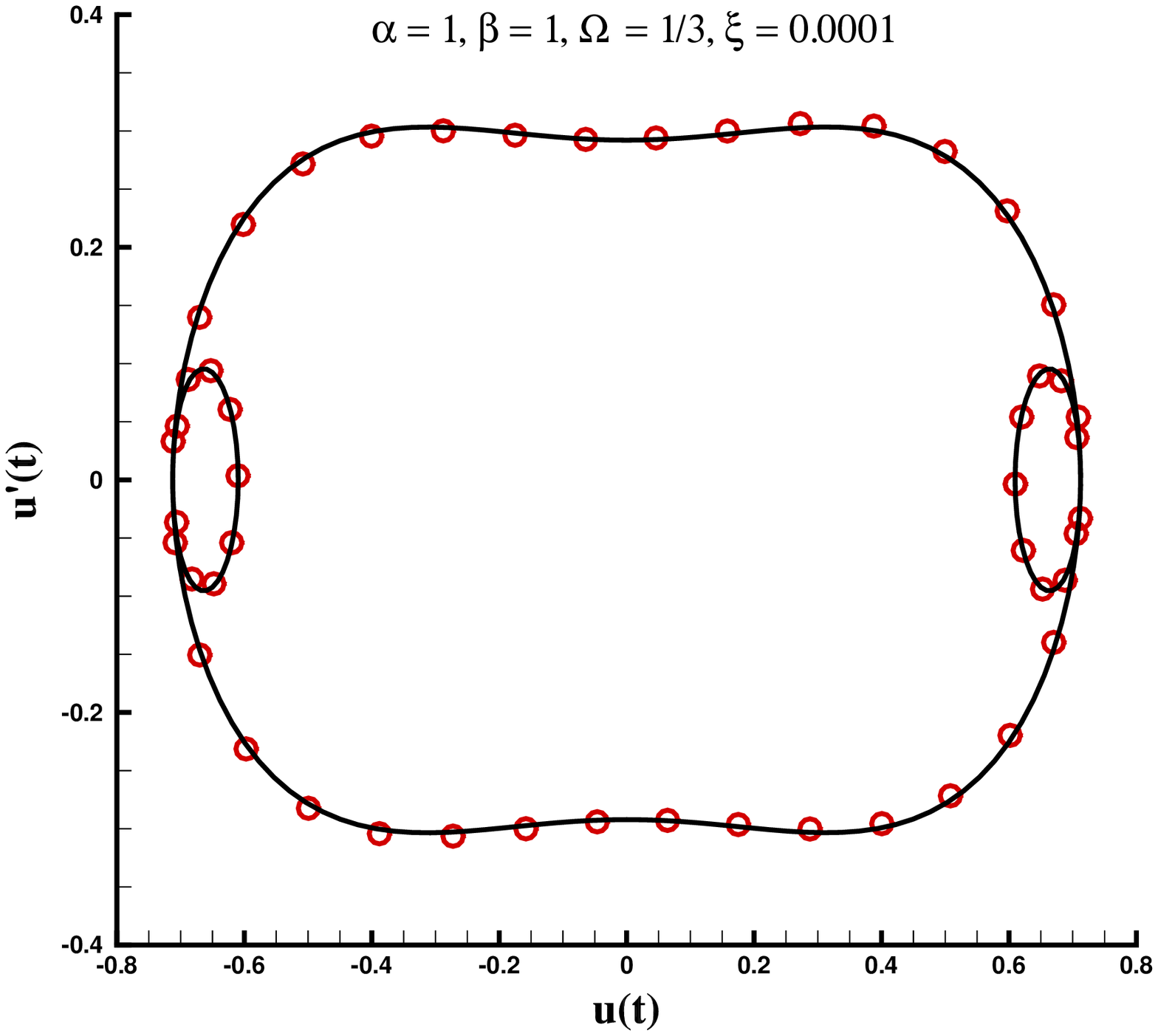}}
             \subfigure[]{\includegraphics[width=2.5in]{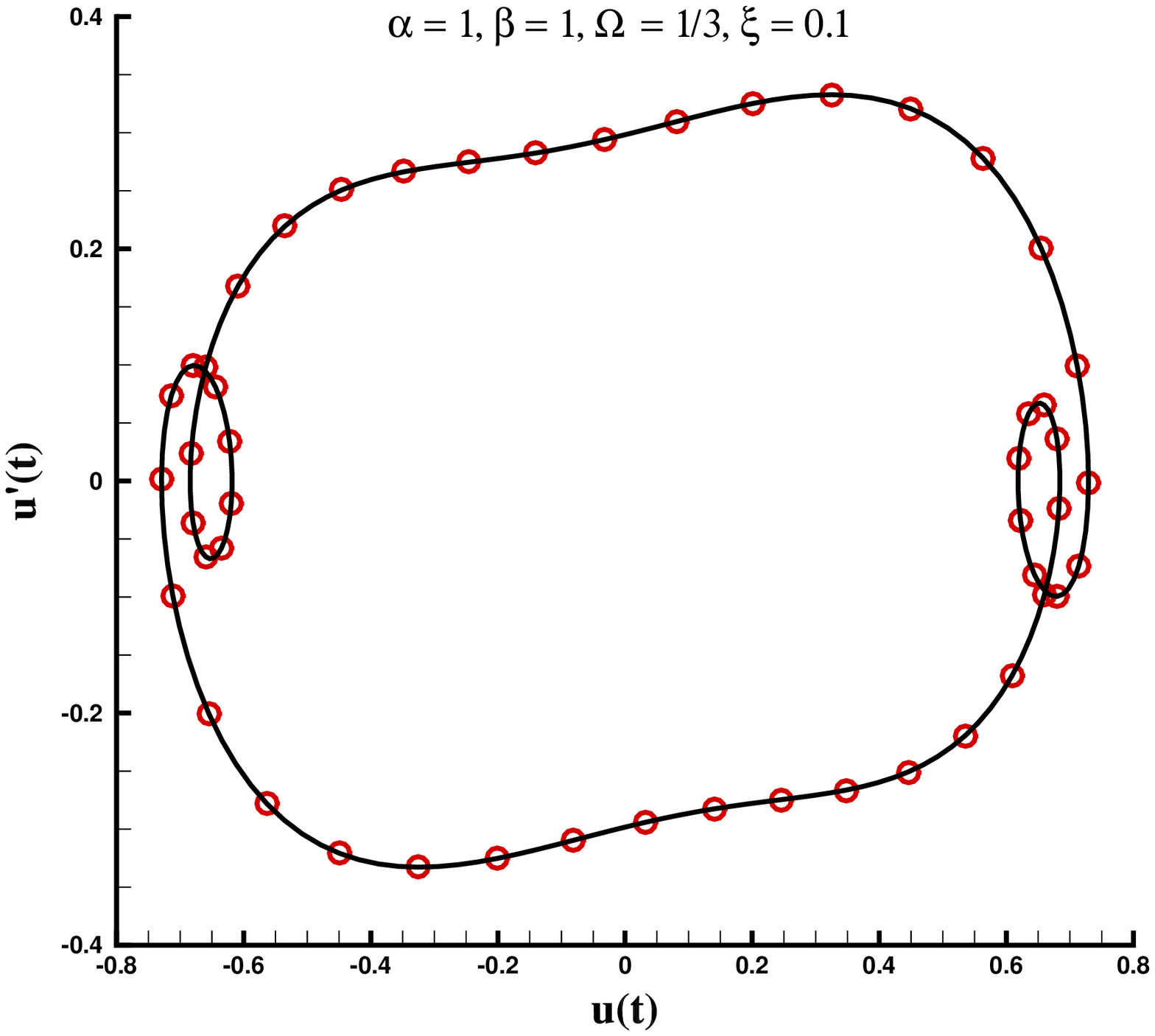}}
        \end{tabular}
 \end{center}
 \caption{Convergent series solutions of  limit-cycle of the forced Duffing equation (\ref{geq:original}) in cases of $\alpha=1, \beta=1,\Omega=1/3$ and different values of $\xi$. Symbols: 10th-order approximation;  Solid line: 30th-order approximation.  (a)  $\xi=10^{-4}$ using $c_{0}=-0.9$ and the initial guess (\ref{initial:A}); (b) $\xi=1/10$ using $c_{0}=-0.8$.  }
  \label{fig:limit-cycle:A}
\end{figure}

  \begin{figure}[tbh]
    \begin{center}
        \begin{tabular}{cc}
             \subfigure[]{\includegraphics[width=2.5in]{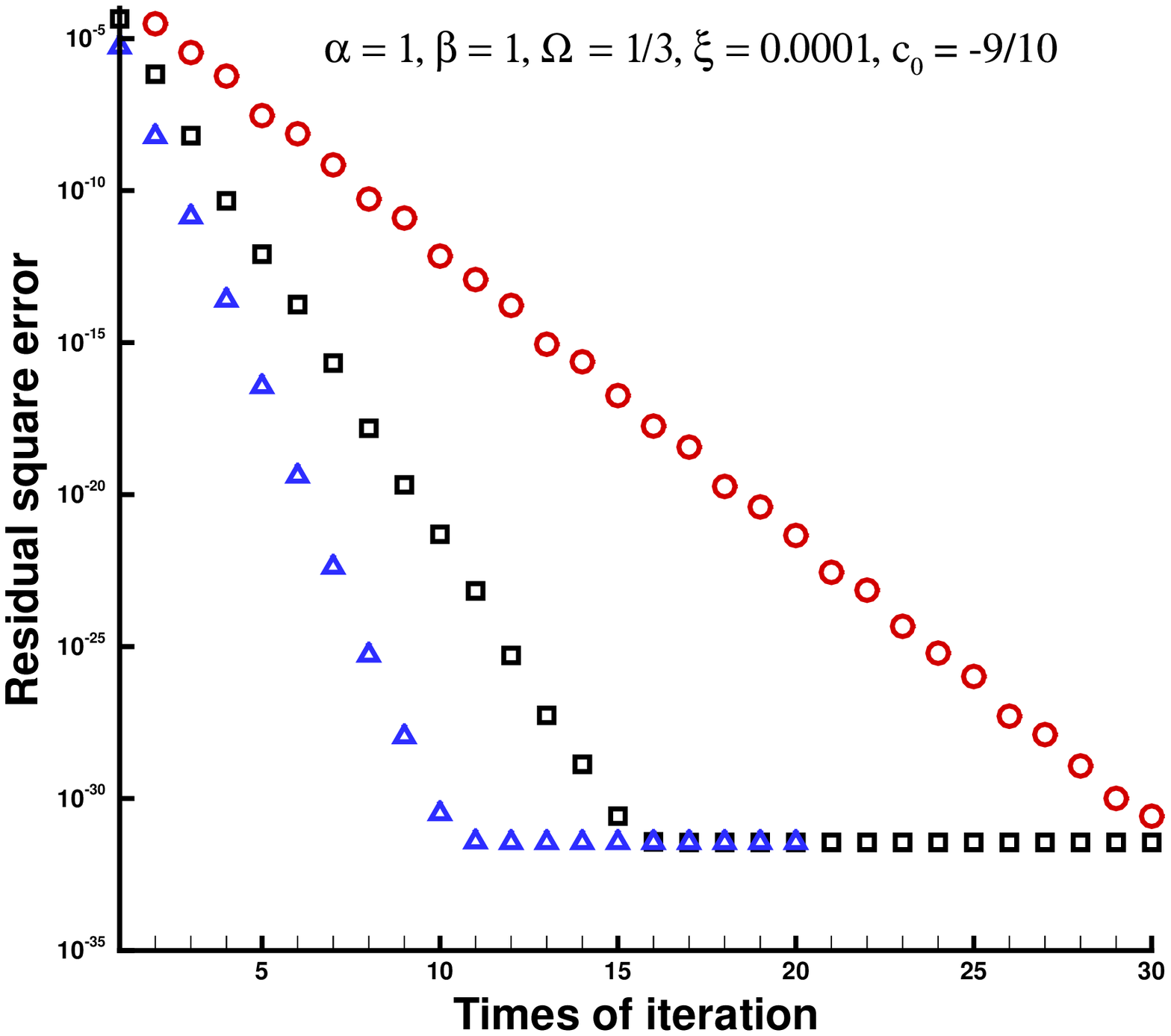}}
             \subfigure[]{\includegraphics[width=2.5in]{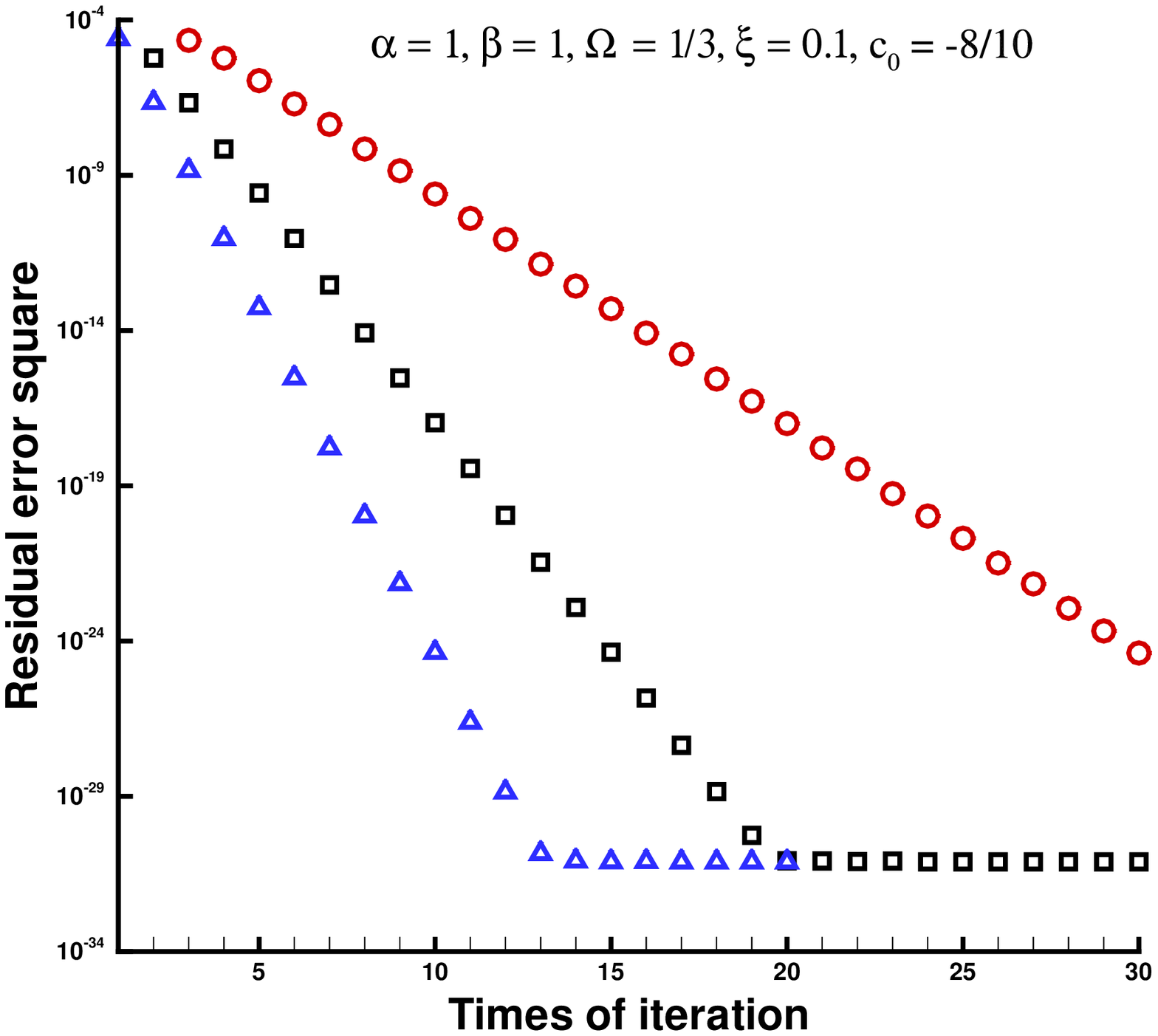}}
        \end{tabular}
 \end{center}
\caption{Residual error square of  the forced Duffing equation (\ref{geq:original}) versus the times of iteration  in case of $\alpha=1, \beta=1$ and $\Omega=1/3$, given by the HAM iteration approach in the case of $\lambda=\sqrt{2}$ and $\delta = 0$ described in \S~4.1.  Cycle: the 1st-order HAM iteration; Square: the 2nd-order HAM iteration; Delta: the 3rd-order HAM iteration. (a) $\xi=0.0001$ using $c_{0} = -0.9$ and the initial guess (\ref{initial:A}); (b) $\xi = 0.1$ using $c_{0 }= -0.8$. }
 \label{fig:RMSi-A}
 \end{figure}

In case of $\lambda=\sqrt{2}$, according to the definitions (\ref{L:inverse}) and (\ref{def:omega[n]}),  we have the  denominator
\begin{equation}
\lambda^{2}-\omega_{n}^{2} = 2-\frac{(2n-1)^{2}}{9} =\frac{17+4n-4n^{2}}{9}.
\end{equation}
Setting $|\lambda^{2}-\omega_{n}^{2} |=\delta=0$, we have the two solutions $n \approx 2.62132$ and $n \approx -1.62132$, which however are {\em not} positive integers.  Thus, nothing belongs to the set $W_{\lambda,\delta}$ when we choose $\lambda=\sqrt{2}$ and $\delta=0$, say, its kernel of the corresponding auxiliary linear operator $\cal L$ has nothing, say, $\ker [{\cal L}] = \emptyset$, corresponding to $W_{\lambda,\delta}=\emptyset$.  Besides, the  corresponding denominators  $(\lambda^{2}-\omega_{n}^{2})$ read
\[  \frac{17}{9}, \;\; 1,\;\;  -\frac{7}{9}, \;\; -\frac{31}{9},\;\; -7, \;\;-\frac{103}{9},  \cdots \]
 which are far away from zero, and therefore {\em all}  denominators are {\bf not} small  so that the ``small  denominator  problem'' does {\bf not} appear at all for {\em arbitrary} values of $\alpha, \beta$ and $\xi$.

 In fact, in the case of $\lambda=\sqrt{2}$ and $\delta=0$, the definitions (\ref{L:inverse}) and (\ref{L:kernel}) are equivalent to such an
 auxiliary linear operator
\begin{equation}
{\cal L}[u] = u'' + 2 \; u  \label{def:L:HAM:1st},
\end{equation}
with the property
\begin{equation}
{\cal L}\Big[A \cos\left(\sqrt{2} \; t\right) + B \sin\left(\sqrt{2} \; t\right)\Big] = 0, \label{L:property:1st}
\end{equation}
whose inverse operator reads
\begin{equation}
{\cal L}^{-1}\Big[ A' \cos(\omega_{n} t) + B' \sin(\omega_{n} t)\Big]= \frac{A' \cos(\omega_{n} t)+ B' \sin(\omega_{n} t)}{2-\omega_{n}^{2}},    \hspace{0.5cm} n\geq 1, \label{def:L:inverse:1st}
\end{equation}
where  $A, B, A', B'$ are arbitrary constants.   Note that (\ref{def:L:HAM:1st}) has no relationship with the original linear operator ${\cal L}_{0}$ defined by (\ref{L:pert}).

In general,  letting $n \geq 1$ be an integer, one can choose $\lambda = 2n \Omega$, which is far away from {\em all} $\omega_{n} = (2n-1)\Omega$, since $ \lambda - \omega_{n}   = \omega_{n+1}-\lambda = \Omega$.  So, in theory there are an infinite number of ways to choose a proper $\lambda$ (and  $\delta=0$) so that  the so-called ``small  denominator  problem'' {\em never} appears for the forced Duffing equation (\ref{geq:original})!

Note that the solution of the $m$th-order deformation equation (\ref{geq:mth}) does not contain the terms $\cos\left(\sqrt{2} \; t\right)$ and $\sin\left(\sqrt{2} \; t\right)$,  since they  do not agree with the solution expression (\ref{solution-expression}) and thus must be disappeared.   This agrees with the conclusion that the kernel of the corresponding auxiliary linear operator $\cal L$ is an empty set, say, $\ker [{\cal L}] = \emptyset$.   Thus, the solution of the $m$th-order deformation equation (\ref{geq:mth}) reads
\begin{equation}
u_{m}(t) = \chi_{m} \; u_{m-1}(t) + c_{0} \; {\cal L}^{-1} \left[ R_{m-1}(t)\right], \hspace{1.0cm} m \geq 1.
\end{equation}

 According to (\ref{def:u[0]}), since $\ker [{\cal L}] = \emptyset$, we choose the initial guess in the form
 \begin{equation}
 u_{0}(t) = \sum_{n=1}^{2} \left[ a_{0,n}\cos(\omega_{n}t) + b_{0,n} \sin(\omega_{n}t) \right], \label{u[0]:A}
 \end{equation}
where the four unknown constants $a_{0,n}, b_{0,n}$ $(n=1,2)$ are determined by enforcing the disappearance of the terms  $\cos(\omega_{1}t)$, $\sin(\omega_{1}t)$, $\cos(\omega_{2}t)$ and $\sin(\omega_{2}t)$ in $R_{0}(t)$.  For details, please refer to \S~3.2.

Without loss of generality, let us first consider here the case of $\alpha=1, \beta=1, \Omega=1/3$ and $\xi=10^{-4}$, corresponding to the ``small  denominator  problem'' from the  viewpoint of perturbation method.
Following the method described in \S~3.2, we have its corresponding initial guess
\begin{eqnarray}
u_{0}(t) &=& 0.775251 \cos(\omega_{1}t) -0.127485 \cos(\omega_{2} t) \nonumber\\
&+ & 4.98191 \times 10^{-5}\sin(\omega_{1} t)-5.24821 \times 10^{-5}\sin(\omega_{2} t),  \label{initial:A}
\end{eqnarray}
which is the {\em unique} real solution of the related set of nonlinear algebraic equations.

Although all physical parameters are given, we always have one unknown parameter, i.e. the convergence-control parameter $c_{0}$, which has no physical means  but can guarantee the convergence of solution series given by the HAM.  To choose an optimal value of $c_{0}$, we check the residual error squares of the first several orders of approximation, defined by (\ref{def:RMS}),  as shown in Figure~\ref{fig:RMS-A}, which give us the optimal value $c_{0}\approx -0.9$ for the considered case.  It is found that, using $c_{0} = -0.9$,  the corresponding solution series indeed converge very quickly: the residual error square decreases about 20 orders of magnitude at the 30th-order of approximation, say, from $1.2\times 10^{-3}$ at the very beginning to $4.1 \times 10^{-23}$, as shown in Table~\ref{table:RMS:A}.  In fact, the 10th-order approximation already agrees quite well with the 30th-order approximation, as shown in Figure~\ref{fig:limit-cycle:A}.  Similarly, we also check the validity of our HAM approach in the case of $\xi=0$, 0.01 and $0.1$, respectively, with $\alpha=1, \Omega=1/3$ and $\beta=1$.  In all of these cases, the solution series converge rather quickly, as shown in Table~\ref{table:RMS:A}.  It is found that, given fixed values of $\alpha, \beta$ and $\Omega$,  the series solutions for small values of $0\leq \xi\leq 0.01$ are almost the same, and the solution series converges  almost at the same rate, as shown in Table~\ref{table:RMS:A}.  This is reasonable in physics,  because the small resistance coefficient $\xi$ has a very small influence on the  limit-cycle.  All of these illustrate the validity of the HAM approach mentioned above.

As mentioned in \S~3.2, one $M$th-order HAM approximation can be used as a new initial guess to gain a better approximation, and so on.  As shown in Figure~\ref{fig:RMSi-A}, in case of $\alpha=1, \beta = 1$, $\Omega=1/3$ and $\xi = 0.0001$,   using the initial guess (\ref{initial:A}), we gain the convergent series solution by means of the first, second and third-order HAM iteration approach, and the corresponding residual error square decreases rather quickly: from $1.2 \times 10^{-3}$ at the beginning to $10^{-23}$ at the 16 iterations for the 2nd-order formula,  or at the 11 iterations for the 3rd-order formula, respectively.  It is found that the higher the order of the HAM iteration approach, the faster the solution series converges.  Similarly, in case of $\xi=0.1$, the iteration also converges rather quickly, as shown in  Figure~\ref{fig:RMSi-A}.   So, choosing an optimal convergence-control parameter $c_{0}$ and using HAM iteration approach,  we can quickly  gain the convergent series solution of the limiting cycle of the forced Duffing equation (\ref{geq:original}) by means of the HAM approach described in \S~2 and \S~3.  This illustrates the validity of our HAM iteration approach.

As shown in  Figure~\ref{fig:RMSi-A},  the residual error square  stops decreasing  at the order of magnitude $10^{-23}$.  This is mainly because, in the solution expression (\ref{solution-expression}) of the limiting cycle $u(t)$, we delete all terms $a_{n}  \cos(\omega_{n}t)$ and $b_{n}  \sin(\omega_{n}t)$ when $|a_{n}|<\epsilon$ and $|b_{n}| <\epsilon$, where we choose $\epsilon = 10^{-20}$ in this paper.  This is specially necessary for the HAM iteration approach, otherwise the number of the base-functions  increases exponentially so that the HAM iteration approach can not work.    It is found that, when a smaller value of  $\epsilon$ such as $\epsilon = 10^{-30}$ is used,  the residual error square  stops  decreasing  at a much smaller level.   However,  for the problem considered in this paper,    $\epsilon = 10^{-20}$  is small enough for the cases under consideration.

Note that, from the viewpoint of perturbation method, the ``small  denominator  problem'' occurs when $\alpha=1, \beta=1$, $\Omega=1/3$ and $ 0.0001\leq \xi \leq 0.01$.   However, as shown in Table~\ref{table:RMS:A},  the series solutions  given by our HAM approach   converge almost in the same rate in case of $0 \leq \xi \leq 0.01$.  Therefore, our HAM approach indeed can  {\bf avoid} the small  denominators.   In other words, from the viewpoint of the HAM approach (in the case of $\lambda=2$ and $\delta=0$), the so-called  ``small  denominator  problem''  does {\bf not} really exist at all: they are just the artifacts of perturbation methods.

\subsection{In case of $\lambda =\omega_{1}$ and $\delta = \Big| \omega_{1}^{2}-\omega^{2}_{\kappa} \Big|$ with $\kappa\geq 1$\label{case:B}}

\begin{table}
\renewcommand\arraystretch{1.5}
\tabcolsep 0pt
\caption{Residual error square of $u(t)$ at different order of approximations of the forced Duffing equation  (\ref{geq:original}) in case of $\alpha=1, \beta=1,\Omega=1/3$ and different values of $\xi$, given by the HAM approach described in \S~4.2  when  $\lambda=\omega_{1}^{2}$ and $\delta=\omega_{2}^{2}-\omega_{1}^{2}$.  }
\vspace*{-7pt}\label{table:RMS:B:K=2}
\begin{center}
\def\temptablewidth{1.0\textwidth}
{\rule{\temptablewidth}{1pt}}
\begin{tabular*}{\temptablewidth}{@{\extracolsep{\fill}}ccccc}
Order of  & $\xi = 0  $  & $\xi=10^{-4} $ & $\xi = 0.01$ & $\xi =0.1  $ \\
approximation &  \hspace{0.5cm} $c_0 = -3/2$\hspace{0.5cm}  & \hspace{0.5cm} $ c_0=-3/2$\hspace{0.5cm} & \hspace{0.5cm}$ c_0=-3/2$\hspace{0.5cm} & \hspace{0.5cm} $c_0=-8/5$\hspace{0.5cm} \\ \hline
0  	& 1.2E-3   & 1.2E-3   & 1.2E-3   & 1.1E-3 \\
1	& 4.0E-4	& 4.0E-4	& 4.0E-4	& 4.1E-4 \\
3	& 1.2E-5	& 1.2E-5	& 1.2E-5	& 1.7E-5 \\
5	& 5.2E-7	& 5.2E-7	& 5.2E-7	& 1.3E-6 \\
10	& 1.7E-10	& 1.7E-10	& 1.7E-10	& 1.4E-9 \\
15	& 1.3E-13	& 1.3E-13 & 1.3E-13	& 2.0E-12 \\
20	& 6.4E-17	& 6.4E-17	& 6.4E-17	& 3.6E-15 \\
25	& 1.2E-19	& 1.2E-19	& 1.2E-19	& 7.6E-18 \\
30	& 7.9E-23	& 7.9E-23 & 7.9E-23 & 1.7E-20 \\		
\end{tabular*}
{\rule{\temptablewidth}{1pt}}
\end{center}
\end{table}

  \begin{figure}[th]
    \begin{center}
        \begin{tabular}{cc}
             \subfigure[]{\includegraphics[width=2.5in]{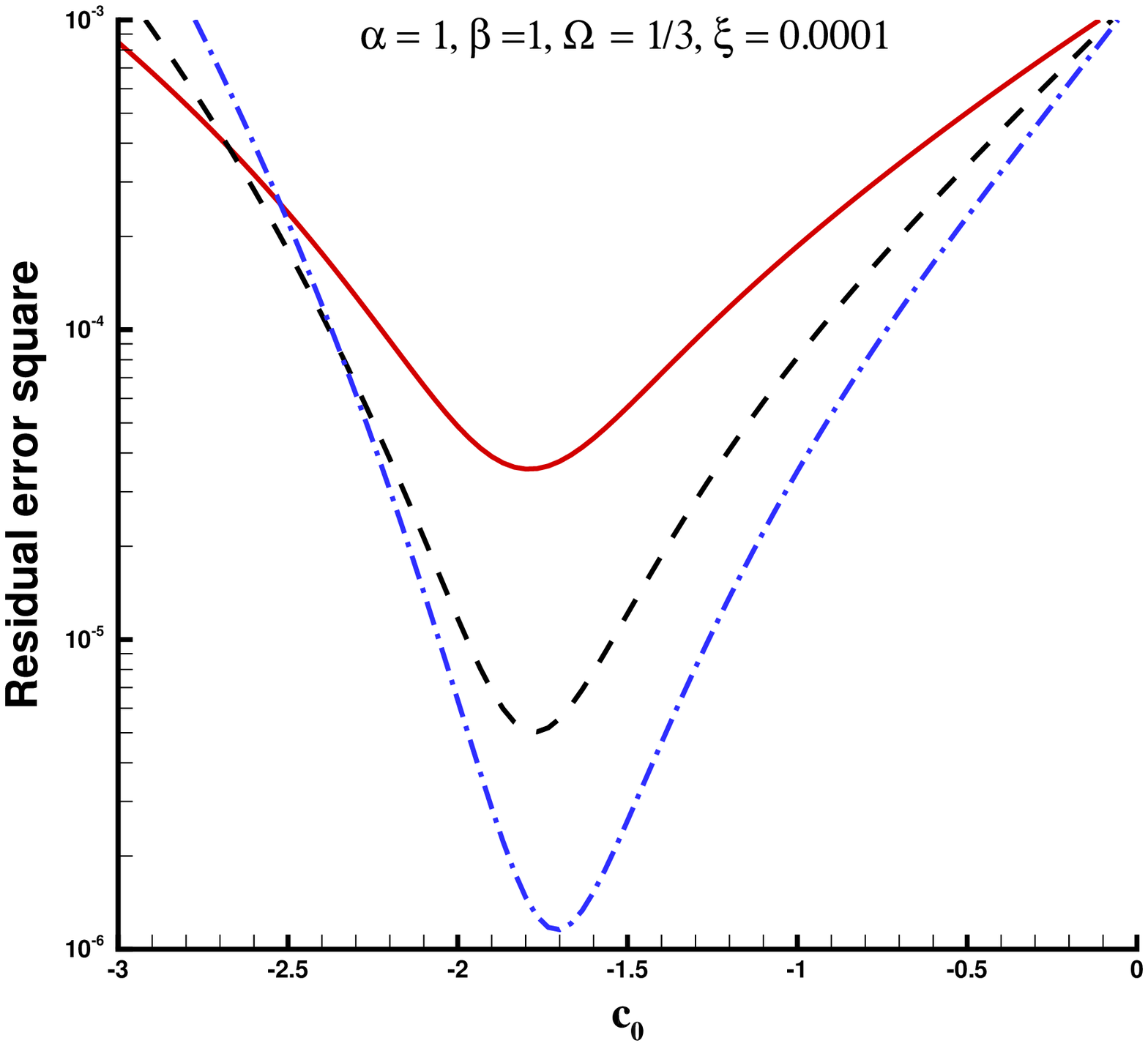}}
             \subfigure[]{\includegraphics[width=2.5in]{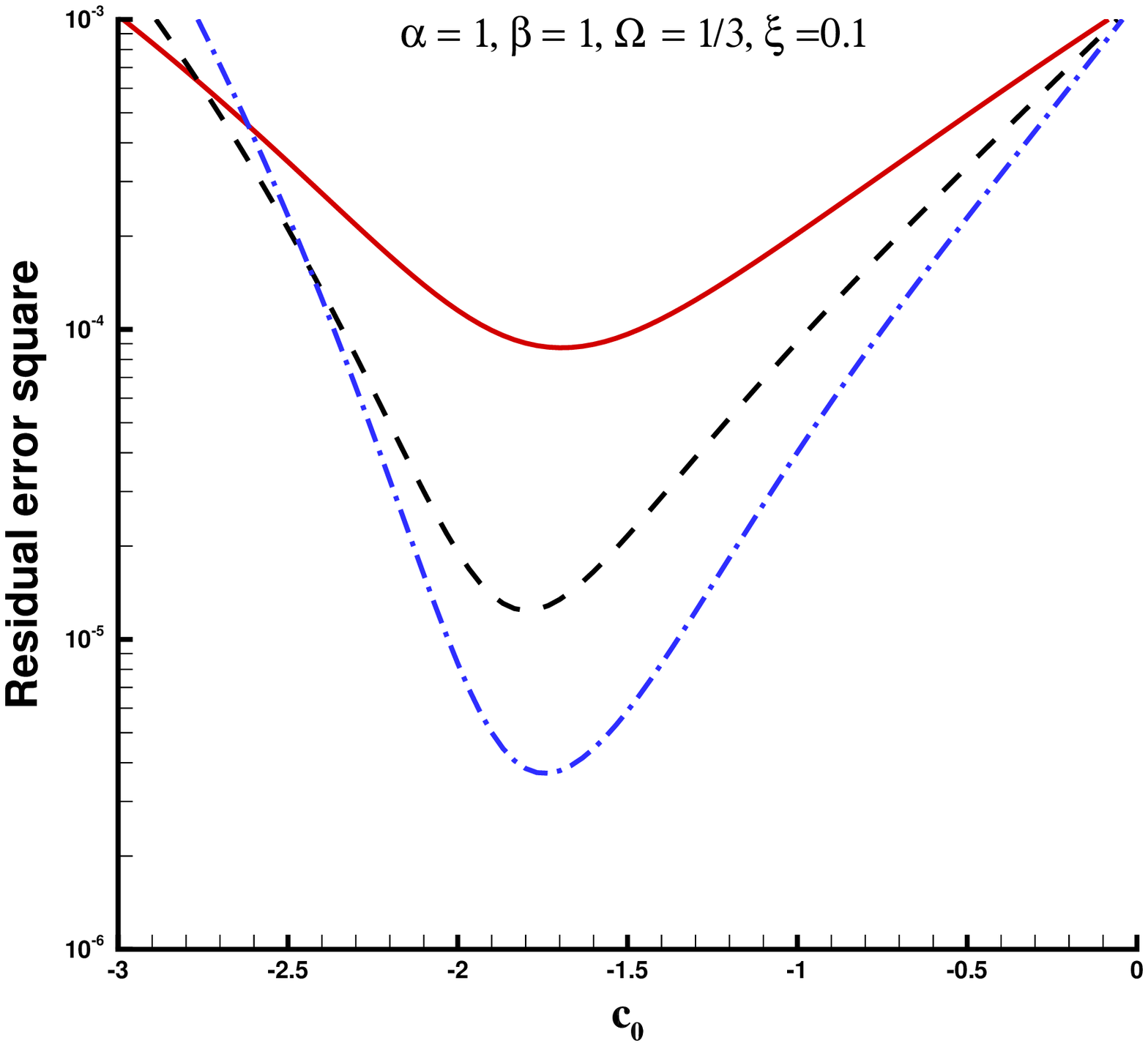}}
        \end{tabular}
 \end{center}
\caption{Residual error squares of  approximations  of $u(t)$ at different-orders versus the convergence-control parameter $c_0$ in case of $\alpha=1, \beta=1, \Omega=1/3$ with different values of $\xi$, given by the HAM approach described in \S~4.2 when $\lambda=\omega_{1}$ and $\delta=\omega_{2}^{2}-\omega_{1}^{2}$.  Solid line: 2nd-order approximation; Dashed-line: 3rd-order approximation; Dash-dotted line: 4th-order approximation.  (a) $\xi =10^{-4}$ using the initial guess (\ref{initial:A});  (b) $\xi=0.1$.  }
\label{fig:RMS:B:K=2}
 \end{figure}

  \begin{figure}[th]
    \begin{center}
             \subfigure[]{\includegraphics[width=2.5in]{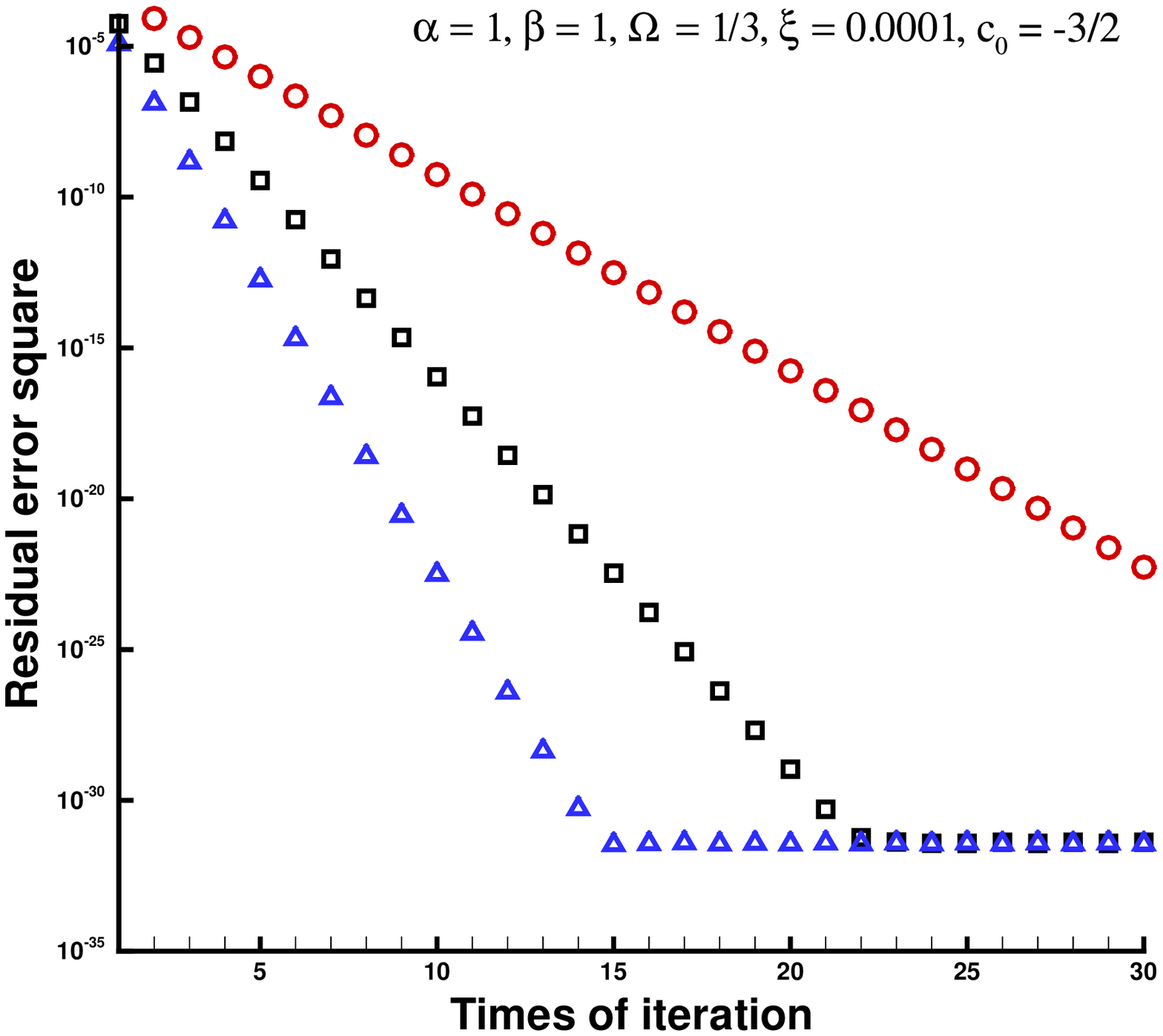}}
             \subfigure[]{\includegraphics[width=2.5in]{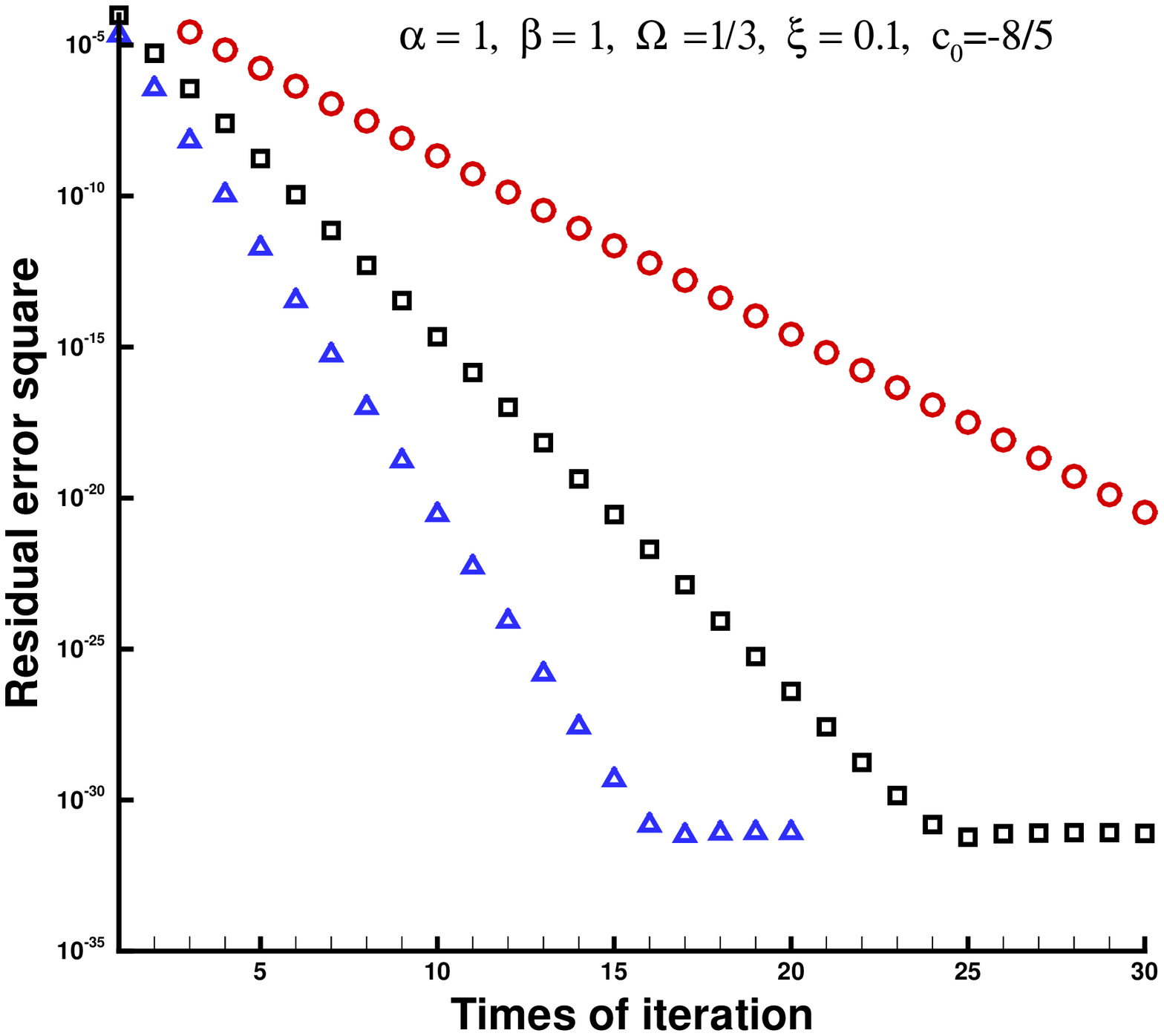}}
 \end{center}
\caption{Residual error square of  the forced Duffing equation (\ref{geq:original}) versus the times of iteration  in case of $\alpha=1, \beta=1, \Omega=1/3$, given by the HAM approach described in \S~4.2 when $\lambda=\omega_{1}$ and $\delta=\omega_{2}^{2}-\omega_{1}^{2}$.  Cycle:  the 2nd-order HAM iteration; Square: the 3rd-order HAM iteration; Delta: the 4th-order HAM iteration.  (a) $\xi=0.0001$  using  the initial guess (\ref{initial:A}) and $c_0=-3/2$; (b)  $\xi = 0.1$ using $c_0=-8/5$.}
 \label{fig:iRMS:B:K=2}
 \end{figure}

Let us further consider the case  $\lambda =\omega_{1}$ and $\delta = \Big| \omega_{1}^{2}-\omega^{2}_{\kappa} \Big|= \omega^{2}_{\kappa}-\omega_{1}^{2}$,  where  $\kappa\geq 1$ is an integer.   According to (\ref{L:kernel}), $|\lambda^{2} - \omega_{n}^{2}|\leq \delta$ leads to the following equation
\begin{equation}
\omega_{n}^{2}-\omega_{1}^{2} \leq  \omega^{2}_{\kappa} - \omega_{1}^{2}, \hspace{1.0cm} n \geq 1,
\end{equation}
which holds for $1\leq n\leq \kappa$.  Thus, the corresponding set $W_{\lambda,\delta}$ has $\kappa$ members, say,
\begin{equation}
W_{\lambda,\delta} = \Big\{\omega_{1}, \omega_{2}, \cdots, \omega_{\kappa} \Big\}.
\end{equation}
In this case, (\ref{L:inverse}) and (\ref{L:kernel}) are equivalent to the following definitions:
\begin{equation}
 {\cal L}^{-1}\Big[ A \cos(\omega_{n} t) + B \sin (\omega_{n} t) \Big] = \frac{A \cos(\omega_{n} t) + B \sin (\omega_{n} t)}{\omega_{1}^{2}-\omega_{n}^{2}}, \hspace{0.5cm}   n > \kappa,  \label{L:inverse:2nd:C}
 \end{equation}
 and
  \begin{equation}
 {\cal L}\left\{  \sum_{n=1}^{\kappa}\Big[A'_{m,n} \cos(\omega_{n} t)+B'_{m,n} \sin(\omega_{n} t)\Big] \right\} = 0,  \label{L:property:2nd:C}
 \end{equation}
 for arbitrary constants $A, B, A'_{m,n}, B'_{m,n}$, where we have great freedom to choose the value of $\kappa$.  According to (\ref{L:property:2nd:C}), the kernel of the corresponding linear operator is a vector space with $\mu = 2\kappa$ dimension, say,
  \begin{equation}
\ker[ {\cal L}] =  \sum_{n=1}^{\kappa}\Big[A'_{m,n} \cos(\omega_{n} t)+B'_{m,n} \sin(\omega_{n} t)\Big].  \label{L:kel:2nd:C}
 \end{equation}
Thus,  in this case, the solution of the $m$th-order  deformation equation (\ref{geq:mth}) reads
 \begin{equation}
 u_m(t)  = \chi_m \; u_{m-1}(t)+c_0 \; {\cal L}^{-1}\Big[R_{m-1}(t)\Big] + \sum_{n=1}^{\kappa}\Big[A'_{m,n} \cos(\omega_{n} t)+B'_{m,n} \sin(\omega_{n} t)\Big], \label{u[m]:2nd:C}
 \end{equation}
 where $A'_{m,n}$ and $B'_{m,n}$ are $2\kappa$ unknown constants, which are determined by enforcing the coefficients of $\cos (\omega_{n} t)$ and $\sin(\omega_{n} t)$ in $R_{m}(t)$ being zero, where $1\leq n\leq \kappa$.

 In this case,  we choose the  initial guess  in the form
 \begin{equation}
 u_{0}(t) = \sum_{n=1}^{\max\{2,\kappa\}} \Big[ a_{0,n} \cos(\omega_{n} t) + b_{0,n} \sin(\omega_{n} t) \Big], \label{u[0]:B}
 \end{equation}
 where the unknown coefficients $a_{0,n}$ and $b_{0,n}$   are determined by enforcing the coefficients of  $\cos (\omega_{n} t)$ and $\sin(\omega_{n} t)$ in $R_{0}(t)$ being zero, where $1 \leq  n  \leq \max\{2, \kappa\}$.

It should be emphasized here that  $\kappa$ can be  greater than 1, since we have great {\em freedom} to choose its value!  For example, when $\kappa=2$, $\ker[{\cal L}]$, i.e. the kernel of the corresponding auxiliary linear operator ${\cal L}$,  is a vector space of {\bf 4} dimension.  When $\kappa=3$, $\ker[{\cal L}]$ is a vector space of  {\bf 6} dimension!  Note that the forced Duffing equation (\ref{geq:original}) is just a 2nd-order nonlinear differential equation.  In the frame of the perturbation method, the forced Duffing equation (\ref{geq:original})  is transferred into an infinite number of 2nd-order linear differential equations, as shown in \S~1.   Note also that, according to the traditional mathemtical theory, the kernel of a second-order linear differential operator is a vector space of {\bf 2} dimension only.  Thus, in case of $\kappa > 1$, our HAM approach is beyond the traditional mathematical theory about differential equations.  This also indicates the novelty of our HAM approach in mathematics.

\subsubsection{Results when $\kappa = 2$}

In this case, we have $\delta = \omega_{2}^{2}-\omega_{1}^{2}$  and that   the kernel of the corresponding auxiliary linear operator $\cal L$ is a vector space of {\bf 4} dimension, say,
\begin{equation}
\ker[{\cal L}] = \sum_{n=1}^{2} \left[  A_{n} \cos(\omega_{n} t) +  B_{n} \sin(\omega_{n} t)  \right],  \label{ker[L]:kappa=2}
\end{equation}
where $A_{n} $ and $B_{n} $ are arbitrary constants.

Without loss of generality, let us consider the case of $\alpha = 1, \beta = 1, \Omega=1/3$ and $\xi=10^{-4}$.
According to (\ref{u[0]:B}), we can choose the same initial guess as (\ref{initial:A}), since they have the same physical parameters.   Note that, unlike all other approximation methods, the HAM contains the so-called ``convergence-control parameter'' $c_{0}$, which has no physical meanings but can guarantee the convergence of the solution series.   As shown in Figure~\ref{fig:RMS:B:K=2}, the optimal convergence-control parameter is about $c_{0}\approx -1.5$.   Using $c_{0}=-3/2$ and the initial guess (\ref{initial:A}), the corresponding series solution converges very quickly, from $1.2 \times 10^{-3}$ at the beginning to   $7.9 \times 10^{-23}$ at the 30th-order of approximation, about 20 orders of magnitude less, as shown in Table~\ref{table:RMS:B:K=2}.   Similarly,  the solution series converge very quickly for $\xi=0, 0.01$ and $0.1$, as shown in Table~\ref{table:RMS:B:K=2}.   It is found that the corresponding limit-cycles given by $\lambda=\omega_{1}$ and $\delta=\omega_{2}^{2}-\omega_{1}^{2}$ in case of $\xi=0.0001$ and $\xi=0.1$ are exactly the same as (a) and (b) in Figure~\ref{fig:limit-cycle:A} given by $\lambda=2$ and $\delta=0$, respectively.
In addition, the 2nd, 3rd and 4th-order HAM iteration formulas also give convergent series solutions rather quickly, as shown in Figure~\ref{fig:iRMS:B:K=2}.  It is found again that, the higher the order of iteration formula, the faster the solution series converges.

All of these confirm the validity of the HAM approach described in \S~4.2.  It is important that, in the case of $\lambda=\omega_{1}$ and $\delta=\omega_{2}^{2}-\omega_{1}^{2}$,  our HAM approach has {\bf nothing} to do with the so-called ``small  denominators''!
In other words, the ``small denominator problem'' never  appears from the viewpoint of the HAM.

Note that, according to traditional mathematical theories, a linear differential operator $\cal L$, whose kernel
is the same as the vector space of {\bf 4} dimension defined by (\ref{ker[L]:kappa=2}), should correspond to the 4th-order differential equation
\begin{equation}
{\cal L}[u] =u^{(4)} + \left( \omega_{1}^{2}+\omega_{2}^{2}\right)u'' + \omega_{1}^{2} \omega_{2}^{2} u = 0, \label{L:kappa=2}
\end{equation}
whose inverse operator reads
\begin{eqnarray}
{\cal L}^{-1}\Big[ A \cos(\omega t) + B \sin(\omega t) \Big] = \frac{A \cos(\omega t) + B \sin(\omega t)}{(\omega^{2}-\omega_{1}^{2})(\omega^{2}-\omega_{2}^{2})},\hspace{0.5cm} \omega \notin\left\{ \omega_{1}, \omega_{2}\right\} \label{L:inverse:kappa=2}
\end{eqnarray}
for arbitrary constants $A$ and $B$.  However,  the above expression is obviously {\em different} from  our inverse operator (\ref{L:inverse:2nd:C}) that  looks like one for a 2nd-order linear differential equation!  In fact, we even do {\em not} know how to explicitly express the corresponding auxiliary linear operator $\cal L$ when $\kappa=2$, say,  $\lambda=\omega_{1}$ and $\delta=\omega_{2}^{2}-\omega_{1}^{2}$ in the HAM approach described in \S~4.2, but fortunately it is {\em unnecessary} to know it in the frame of the HAM!    The most important fact is that our HAM-based approach is valid and the corresponding solution series of the limiting cycle converge quickly, as shown in Table~\ref{table:RMS:B:K=2} and Figures~\ref{fig:RMS:B:K=2} \& \ref{fig:iRMS:B:K=2}.  This verifies the validity and novelty of our HAM approach mentioned above.

\subsubsection{Results when $\kappa = 3$}

In this case we have $\delta = \omega_{3}^{2}-\omega_{1}^{2}$  and  the kernel of the auxiliary linear operator $\cal L$ is a vector space of {\bf 6} dimensions, say,
\begin{equation}
\ker[{\cal L}] = \sum_{n=1}^{3} \left[  A_{n} \cos(\omega_{n} t) +  B_{n} \sin(\omega_{n} t)  \right],  \label{ker[L]:kappa=3}
\end{equation}
where $A_{n} $ and $B_{n} $ are arbitrary constants.

Besides, according to (\ref{u[0]:B}), the initial guess should be in the form
 \begin{equation}
 u_{0}(t) = \sum_{n=1}^{3} \Big[ a_{0,n} \cos(\omega_{n} t) + b_{0,n} \sin(\omega_{n} t) \Big]. \label{u[0]:B:K=3}
 \end{equation}
 Substituting it into the forced Duffing equation (\ref{geq:original}) and enforcing the coefficients of the terms $\cos(\omega_{n}t)$ and $\sin(\omega_{n}t)$ $(n=1,2,3)$ being zero, we have a set of {\em six} nonlinear algebraic equations, whose real solutions determine the  six  unknown constants in (\ref{u[0]:B:K=3}).
  It is interesting that the set of these six nonlinear algebraic equations has multiple real solutions (complex solutions have no physical meanings here)
 in many cases, for example,  such as $\alpha=1, \xi=0, \Omega=1/3$ but a large value of $\beta$, i.e.  $\beta=5$:
 \begin{eqnarray}
&& a_{0,1}=0.333781, a_{0,2}=0.107352, a_{0,3} = -0.509166; \label{u[0]:beta=5:xi=0:a} \\
&&  a_{0,1}=0.526136, a_{0,2}=-0.11976, a_{0,3} = 0.181401; \label{u[0]:beta=5:xi=0:b} \\
&&   a_{0,1}=0.482000, a_{0,2}=0.0264671, a_{0,3} = -0.200815;  \label{u[0]:beta=5:xi=0:c}
 \end{eqnarray}
with $b_{0,1}=b_{0,2}=b_{0,3}=0$, respectively, corresponding to the three initial guesses in the form (\ref{u[0]:B:K=3}).

\begin{figure}[tbh!]
    \begin{center}
        \begin{tabular}{cc}
             \subfigure[]{\includegraphics[width=2.5in]{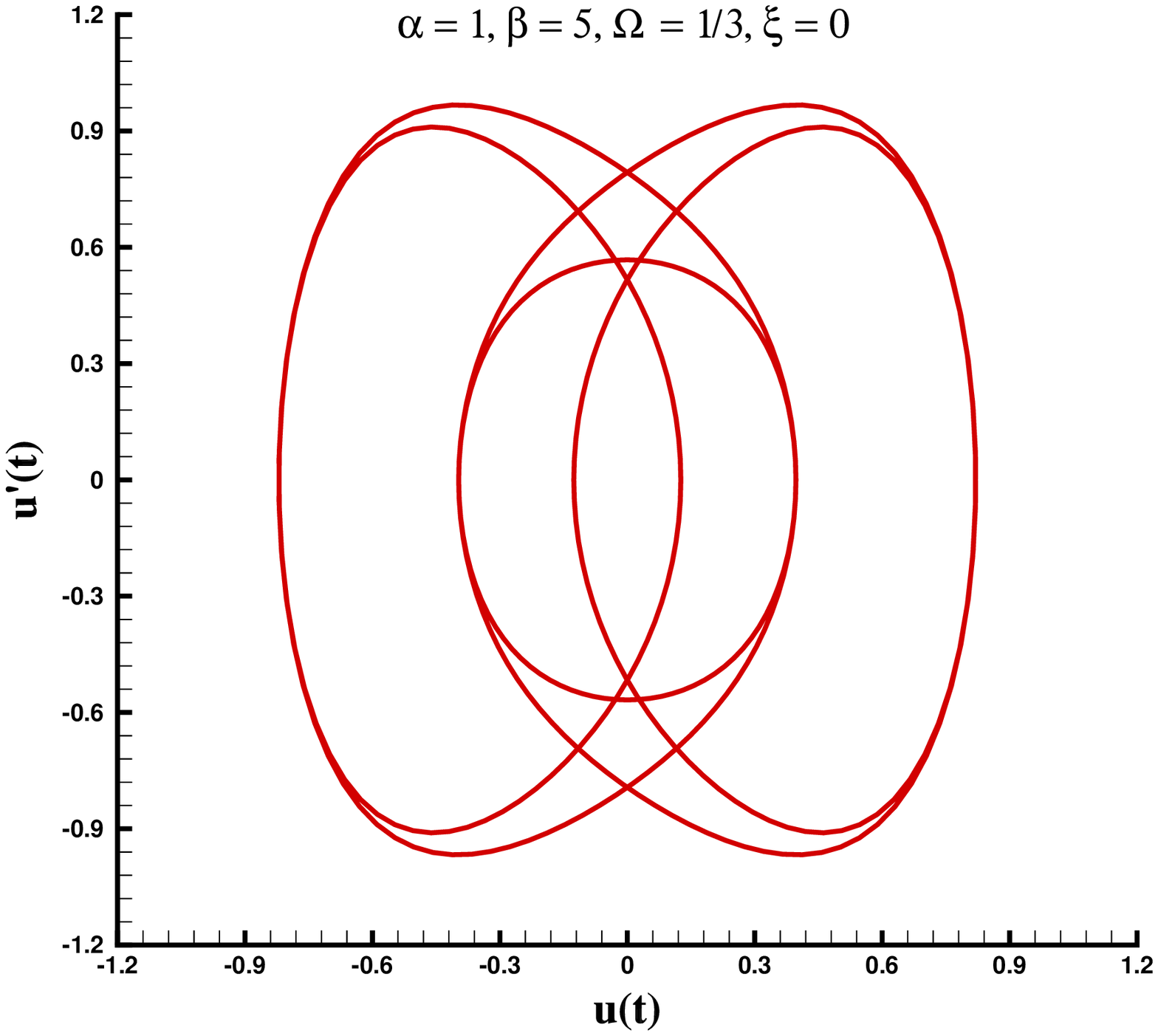}}
             \subfigure[]{\includegraphics[width=2.5in]{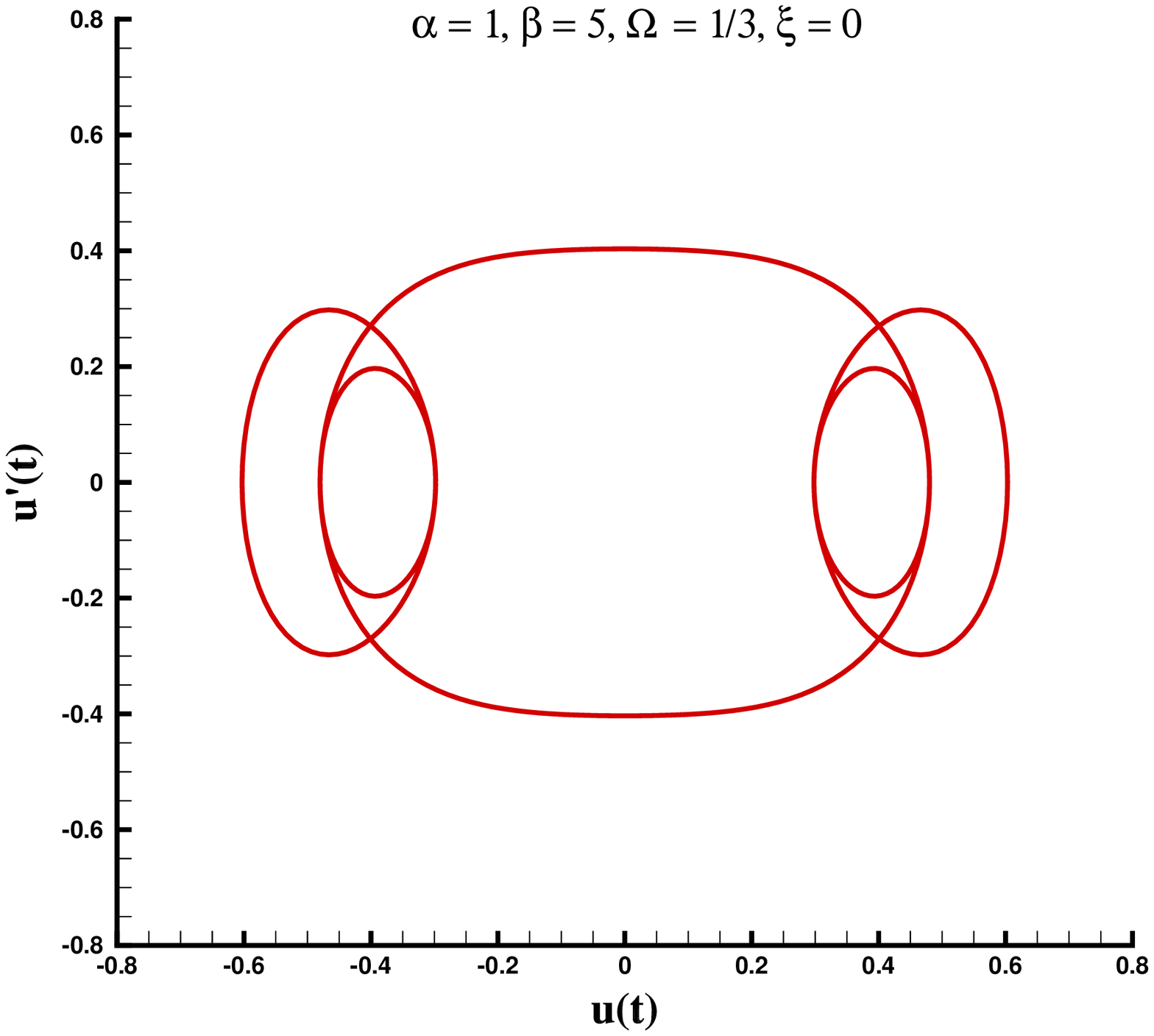}}\\
             \subfigure[]{\includegraphics[width=2.5in]{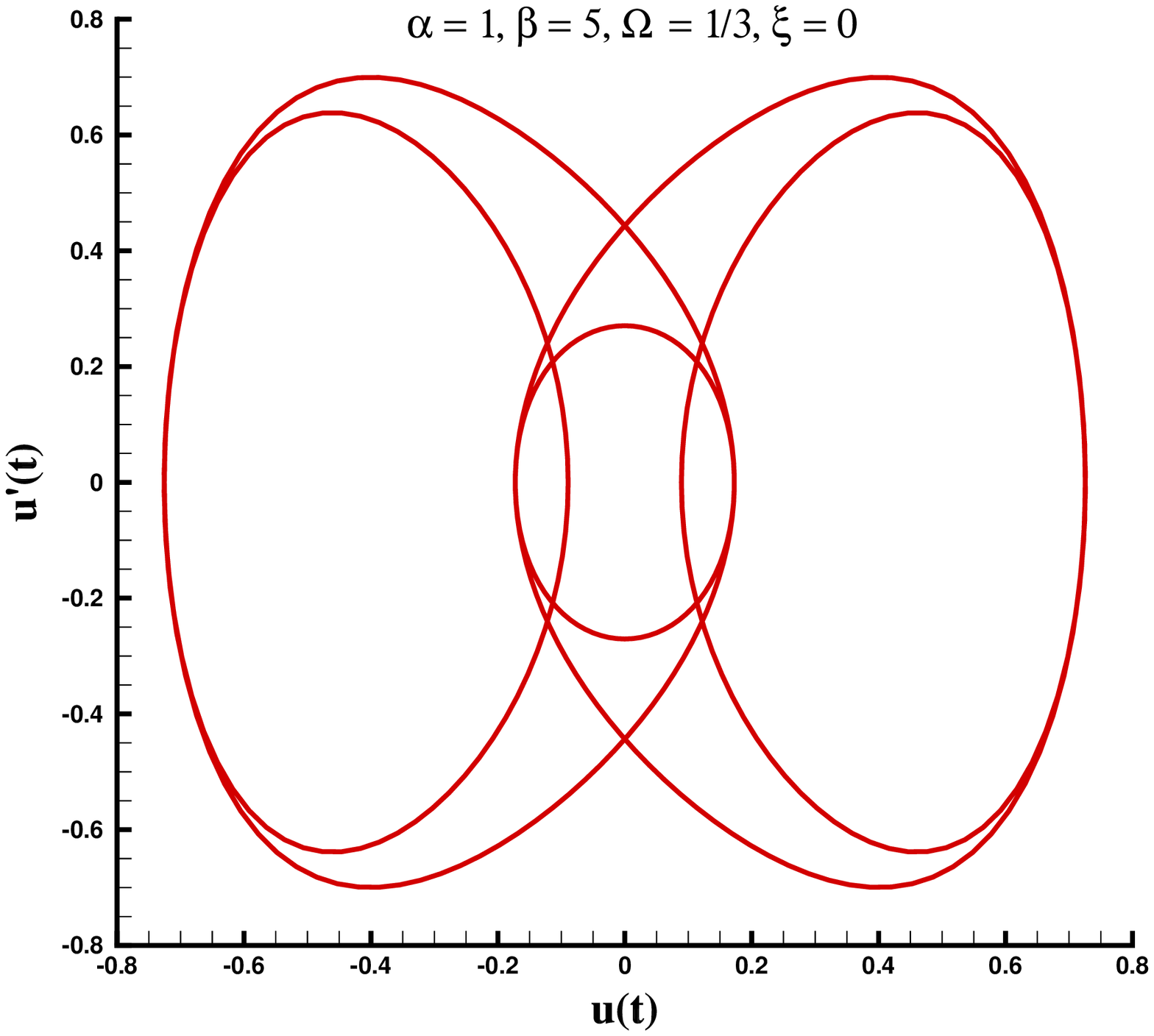}}
             \subfigure[]{\includegraphics[width=2.5in]{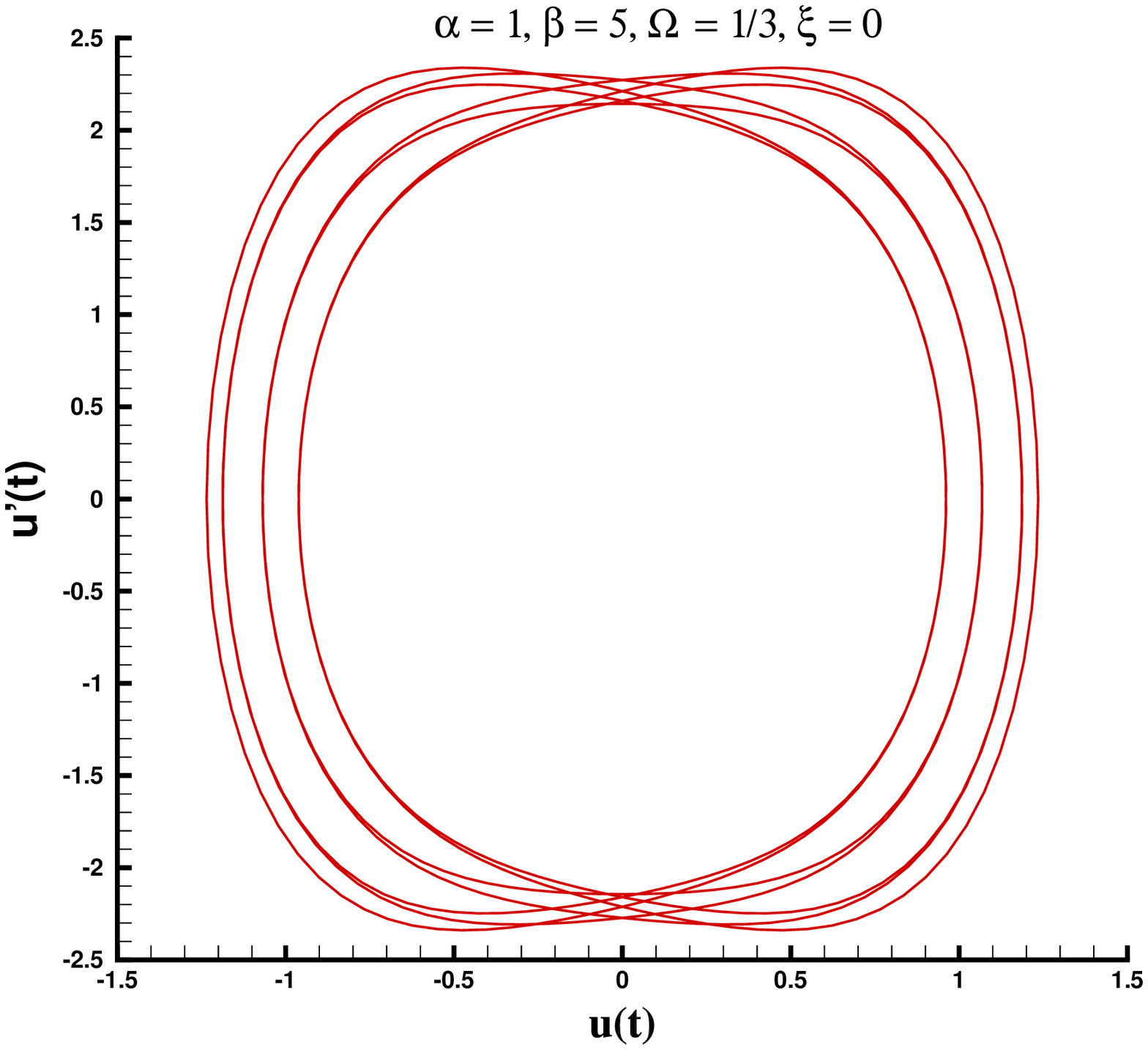}}
        \end{tabular}
 \end{center}
 \caption{Multiple limit-cycles of the forced Duffing equation (\ref{geq:original}) in cases of $\alpha=1,\Omega=1/3, \xi = 0$ and $\beta=5$, given by the 2nd-order HAM iteration described in \S~4.2 when $\lambda=\omega_{1}$ and $\delta=\omega_{\kappa}^{2}-\omega_{1}^{2}$.  (a) using the initial guess  (\ref{u[0]:beta=5:xi=0:a}), $c_{0}=-1$ and $\kappa=3$; (b) using the initial guess  (\ref{u[0]:beta=5:xi=0:b}), $c_{0}=-3/2$ and $\kappa=3$; (c) using the initial guess  (\ref{u[0]:beta=5:xi=0:c}), $c_{0}=-1$ and $\kappa=3$; (d) using the initial guess (\ref{u[0]:beta=5:xi=0:K=4:d}),  $c_{0}=-1$ and $\kappa=4$.  }
 \label{fig:limit-cycle:K3}
\end{figure}

Using the initial guess (\ref{u[0]:beta=5:xi=0:a}) and the corresponding optimal convergence-control parameter $c_{0}=-1$, we gain a convergent  series  solution, shown as (a) in Figure~\ref{fig:limit-cycle:K3},  by means of the 2nd-order HAM iteration: the residual error square of the forced Duffing equation (\ref{geq:original}) decreases from  0.11 at the beginning to  $4.9 \times 10^{-22}$ at the 30th iteration.

  Using the initial guess (\ref{u[0]:beta=5:xi=0:b}) and the corresponding optimal convergence-control parameter $c_{0}=-3/2$, we gain the convergent  series  solution, shown as (b) in Figure~\ref{fig:limit-cycle:K3},  by means of the 2nd-order HAM iteration: the residual error square of the forced Duffing equation (\ref{geq:original}) decreases from  0.11 at the beginning to  $2.1 \times 10^{-30}$ at the 20th iteration.

Using the initial guess (\ref{u[0]:beta=5:xi=0:c}) and the corresponding optimal convergence-control parameter $c_{0}=-1$, we gain the convergent  series  solution, shown as (c) in Figure~\ref{fig:limit-cycle:K3},  by means of the 2nd-order HAM iteration: the residual error square of the forced Duffing equation (\ref{geq:original}) decreases from  0.11 at the beginning to  $7.7 \times 10^{-30}$ at the 30th iteration.

 It is interesting that  we have  the three initial guesses (\ref{u[0]:beta=5:xi=0:a}) - (\ref{u[0]:beta=5:xi=0:c}) in the case of $\kappa=3$, which give us three different  limit-cycles, as shown in (a), (b) and (c) of Figure~\ref{fig:limit-cycle:K3}, respectively.   It should be emphasized that,  it is the HAM that provides us such kind of great freedom to choose the initial guess.  Note also that, in the frame of the perturbation method, there exists the {\em unique} initial guess (\ref{initial:pert}) only, and thus in theory  it is impossible to find these multiple limit-cycles by the perturbation method\footnote{In fact, as mentioned in \S~1, the unique perturbation series diverges even when $\beta=0.012$ in case of $\Omega=1/3$ and $\alpha=1$, corresponding to a very weak nonlinearity}.  This illustrates the advantages and novelty of the HAM beyond perturbation.

\begin{figure}[tbh!]
    \begin{center}
        \begin{tabular}{cc}
             \subfigure[]{\includegraphics[width=2.5in]{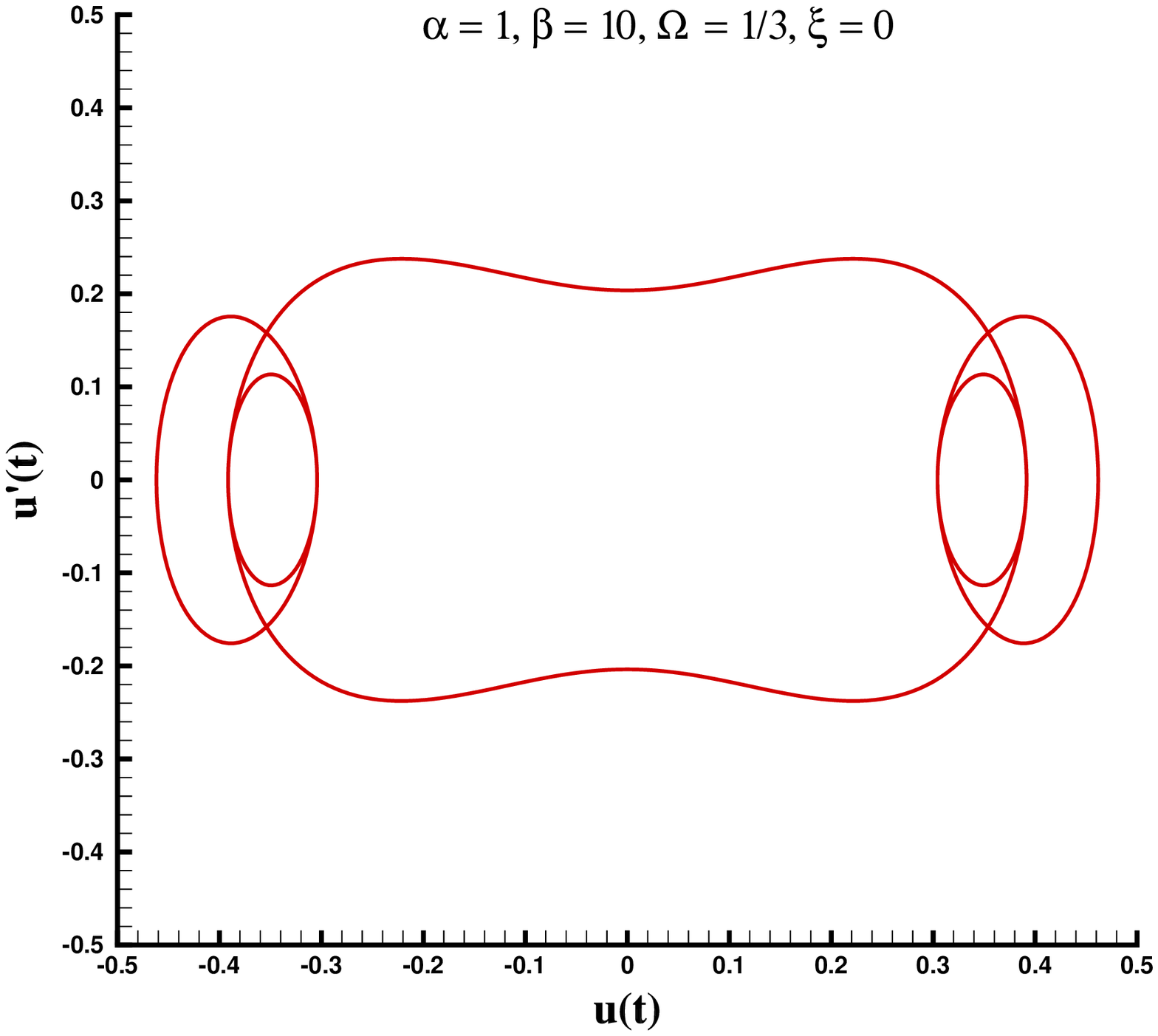}}
             \subfigure[]{\includegraphics[width=2.5in]{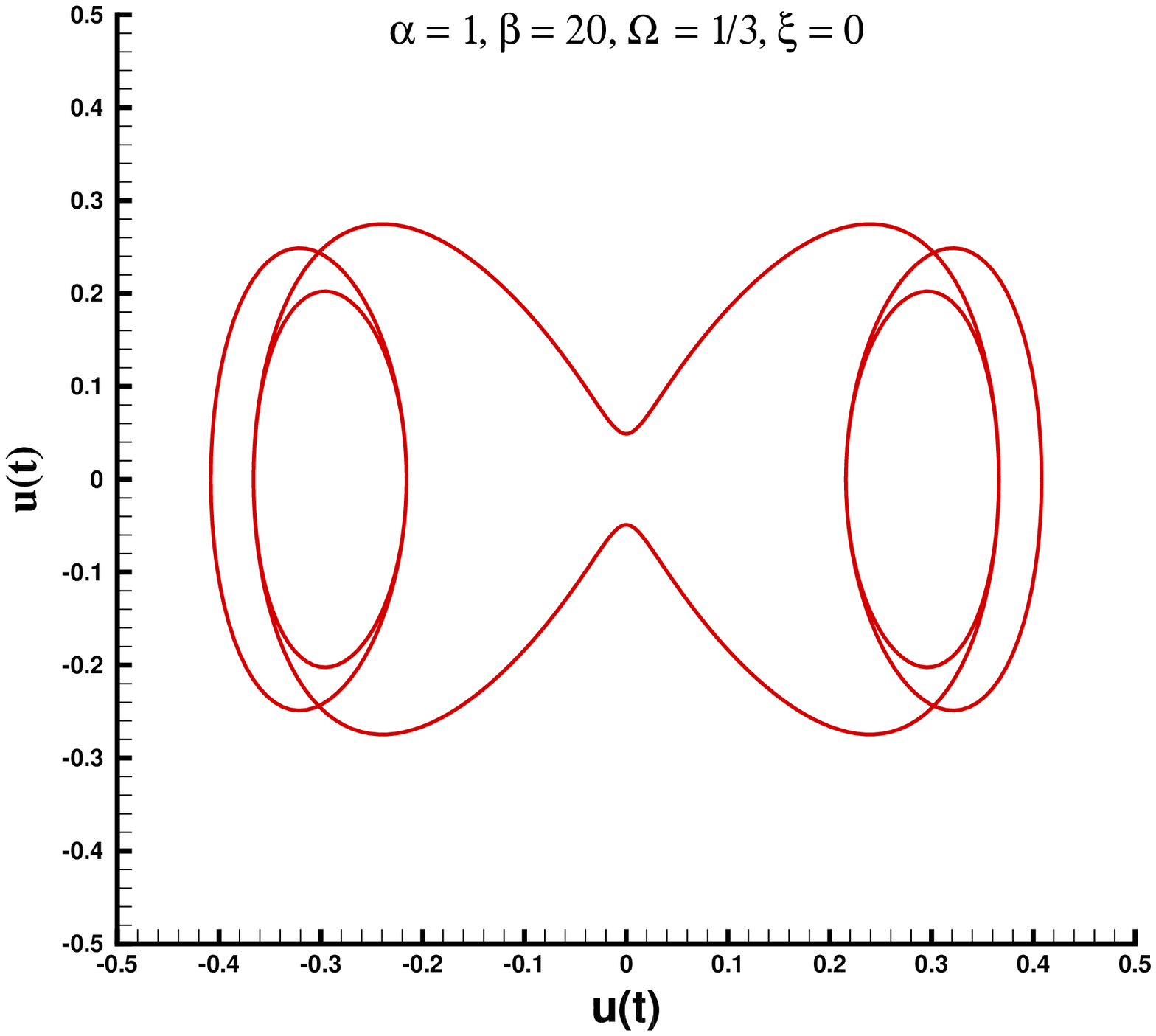}}\\
             \subfigure[]{\includegraphics[width=2.5in]{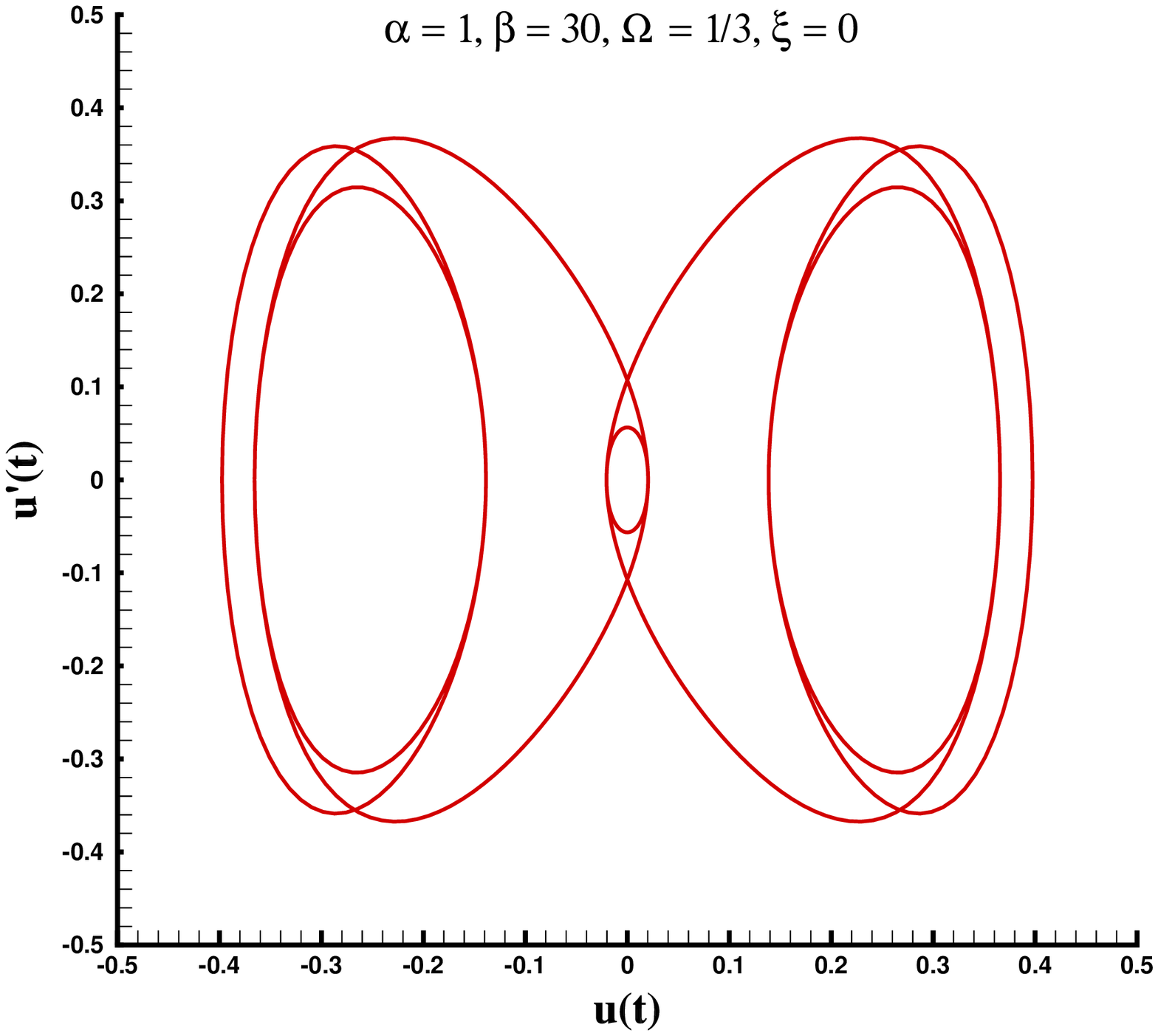}}
             \subfigure[]{\includegraphics[width=2.5in]{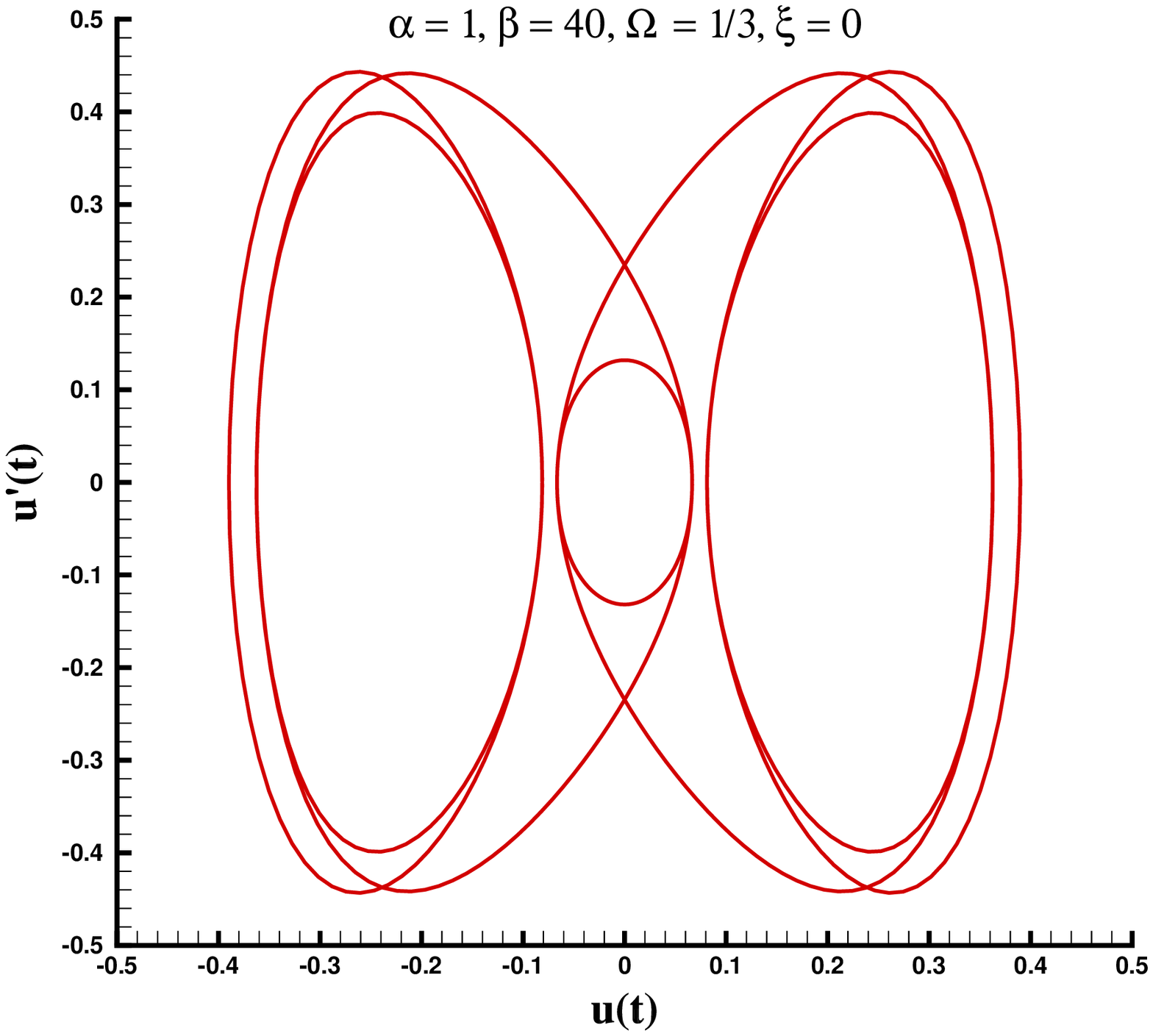}}
        \end{tabular}
 \end{center}
 \caption{Multiple limit-cycles of the forced Duffing equation (\ref{geq:original}) in cases of $\alpha=1,\Omega=1/3, \xi = 0$ and some large values of $\beta$, given by the 2nd-order HAM iteration described in \S~4.2 when $\lambda=\omega_{1}$ and $\delta=\omega_{3}^{2}-\omega_{1}^{2}$, corresponding to $\kappa=3$.  (a) $\beta=10$  using the  unique  initial guess (\ref{u[0]:beta=10:xi=0:K=3})  and   $c_{0}=-3/2$; (b) $\beta=20$  using  the  unique  initial guess (\ref{u[0]:beta=20:xi=0:K=3})  and   $c_{0}=-1/2$; (c) $\beta=30$  using  the  unique  initial guess (\ref{u[0]:beta=30:xi=0:K=3})  and  $c_{0}=-1/5$; (d) $\beta=40$ using the  unique  initial guess (\ref{u[0]:beta=40:xi=0:K=3})  and  $c_{0}=-1/25$.  }
 \label{fig:limit-cycle:K3:high-nonlinearity}
\end{figure}

Note that the forced Duffing equation (\ref{geq:original}) contains the nonlinear term $\beta u^{3}$.  So, the larger the value of $\beta$, the higher the nonlinearity of the Duffing equation.  As mentioned in \S~1, the perturbation approach is invalid even for $\beta \geq 0.012$ and small $\xi$.   However, using our HAM approach in a similar way, we can gain convergent series solution even in the cases with rather high nonlinearity, such as $\alpha=1, \Omega=1/3, \xi=0$ and $10\leq \beta \leq 40$, as shown in Figure~\ref{fig:limit-cycle:K3:high-nonlinearity}.   It is found that,  when $\kappa = 3$,  using the approach mentioned in \S~3.2,  there exist only one initial guess in the form (\ref{u[0]:B:K=3}) for each $\beta \in [10, 40]$, say,
\begin{eqnarray}
&& \beta=10: a_{0,1}= 0.450482, a_{0,2}= -0.0931627, a_{0,3}= 0.0779936; \label{u[0]:beta=10:xi=0:K=3}\\
&& \beta = 20:   a_{0,1}= 0.371919,  a_{0,2}= -0.0737289, a_{0,3}= 0.0477389;  \label{u[0]:beta=20:xi=0:K=3}\\
&& \beta= 30: a_{0,1}= 0.330788, a_{0,2}=-0.0646278, a_{0,3}=0.0380275;  \label{u[0]:beta=30:xi=0:K=3}\\
&& \beta =40: a_{0,1}=0.303900, a_{0,2}=-0.0589144, a_{0,3}=0.0328625, \label{u[0]:beta=40:xi=0:K=3}
\end{eqnarray}
where $b_{0,n} = 0$ for $n=1,2,3$, corresponding to the four initial guesses in the form (\ref{u[0]:B:K=3}).   In all of these cases,  the convergent series solutions are obtained by means of the 2nd-order HAM iteration using a proper convergence-control parameter $c_{0}$, as shown in Figure~\ref{fig:limit-cycle:K3:high-nonlinearity}.  Thus, our HAM  approach is indeed valid for high nonlinearity.    Besides, it should be emphasized that the ``small denominator problem'' never appears in all cases.
This illustrates the validity  of  the  HAM  approach  for high nonlinearity in the case of $\lambda=\omega_{1}$ and $\delta=\omega_{3}^{2}-\omega_{1}^{2}$ and its advantages beyond perturbation.

Note that, according to traditional mathematical theories, a linear differential operator $\cal L$, whose kernel
is the same as the vector space of {\bf 6} dimension defined by (\ref{ker[L]:kappa=3}), should correspond to the 6th-order differential equation
\begin{eqnarray}
{\cal L}[u] &=& u^{(6)} + \left( \omega_{1}^{2}+\omega_{2}^{2}+\omega_{3}^{2}\right)u^{(4)} \nonumber\\
&+&\left(  \omega_{1}^{2} \omega_{2}^{2}+\omega_{1}^{2} \omega_{3}^{2} +\omega_{2}^{2} \omega_{3}^{2}\right) u'' + \omega_{1}^{2} \omega_{2}^{2} \omega_{3}^{2} \; u = 0, \label{L:kappa=3}
\end{eqnarray}
whose inverse operator reads
 \begin{eqnarray}
&& {\cal L}^{-1}\Big[ A \cos(\omega t) + B \sin(\omega t) \Big] \nonumber\\
&=& \frac{A \cos(\omega t) + B \sin(\omega t)}{(\omega^{2}-\omega_{1}^{2})(\omega^{2}-\omega_{2}^{2})(\omega^{2}-\omega_{3}^{2})},  \hspace{0.5cm} \omega \notin\left\{ \omega_{1}, \omega_{2}, \omega_{3}\right\} \label{L:inverse:kappa=3}
\end{eqnarray}
for arbitrary constants $A$ and $B$. However,  the above expression is obviously {\em different} from our inverse operator (\ref{L:inverse:2nd:C}) that looks like one for a 2nd-order linear differential equation!  In fact, we even do {\em not} know how to explicitly express the corresponding auxiliary linear operator $\cal L$ when $\kappa=3$, say,  $\lambda=\omega_{1}$ and $\delta=\omega_{3}^{2}-\omega_{1}^{2}$, but fortunately it is {\em unnecessary} to know it in the frame of the HAM.   The most important fact  is  that our HAM-based approach is valid and the corresponding solution series converge quickly, as mentioned above, which verifies the validity and novelty of our HAM approach.

\subsubsection{Results given by $\kappa = 4$}

    \begin{figure}[tbh!]
    \begin{center}
        \begin{tabular}{cc}
             \subfigure[]{\includegraphics[width=2.5in]{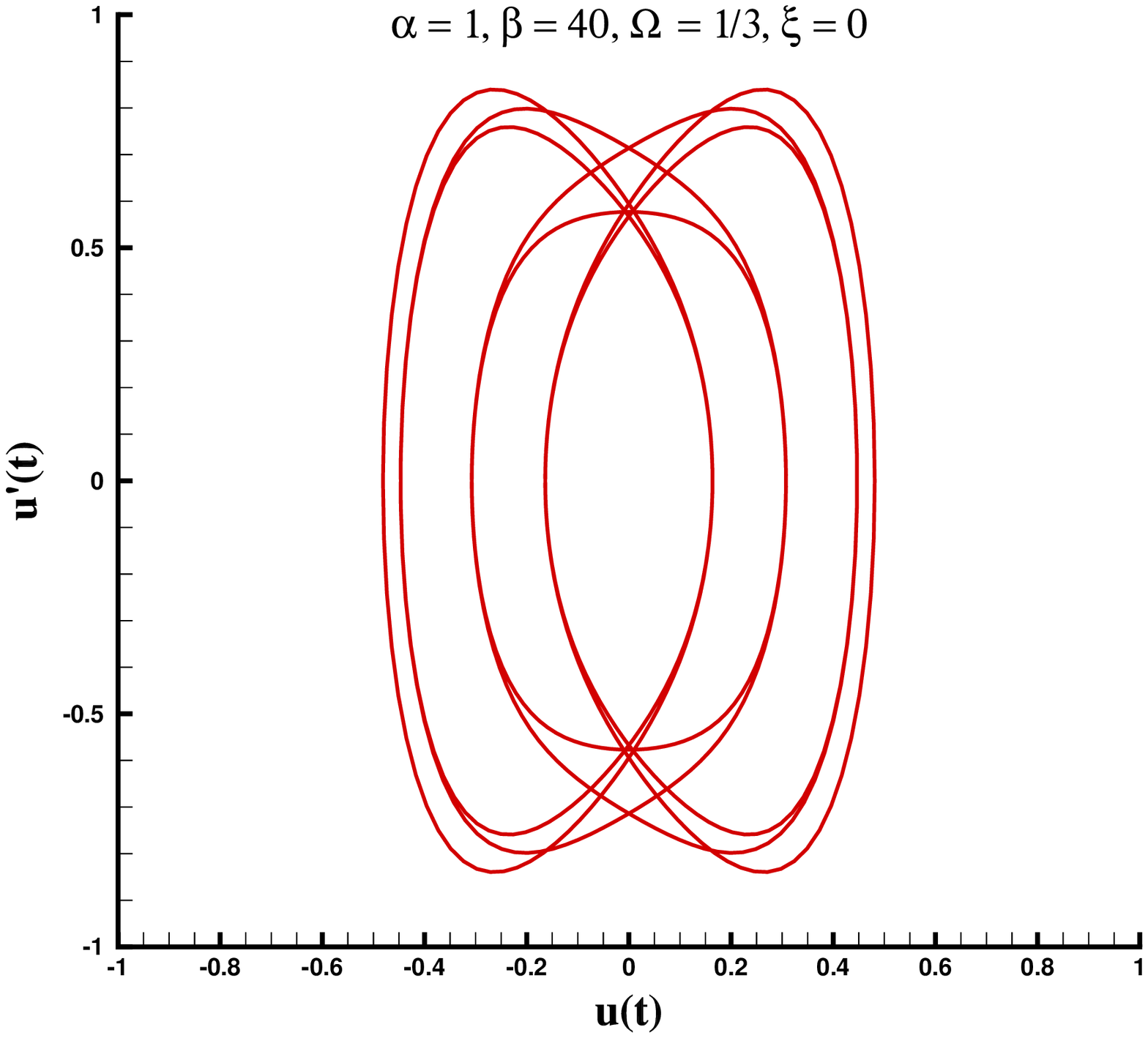}}
             \subfigure[]{\includegraphics[width=2.5in]{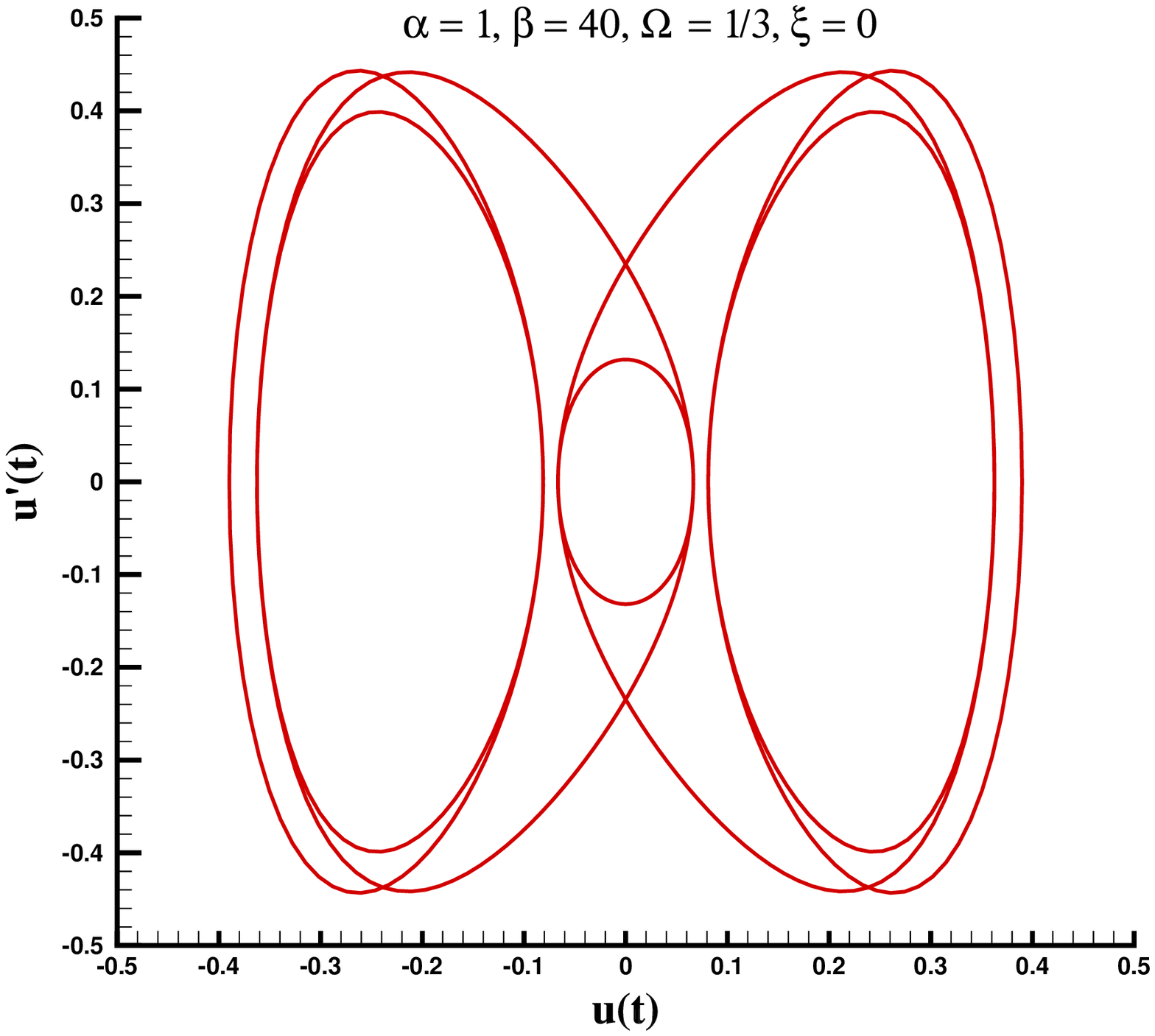}}\\
             \subfigure[]{\includegraphics[width=2.5in]{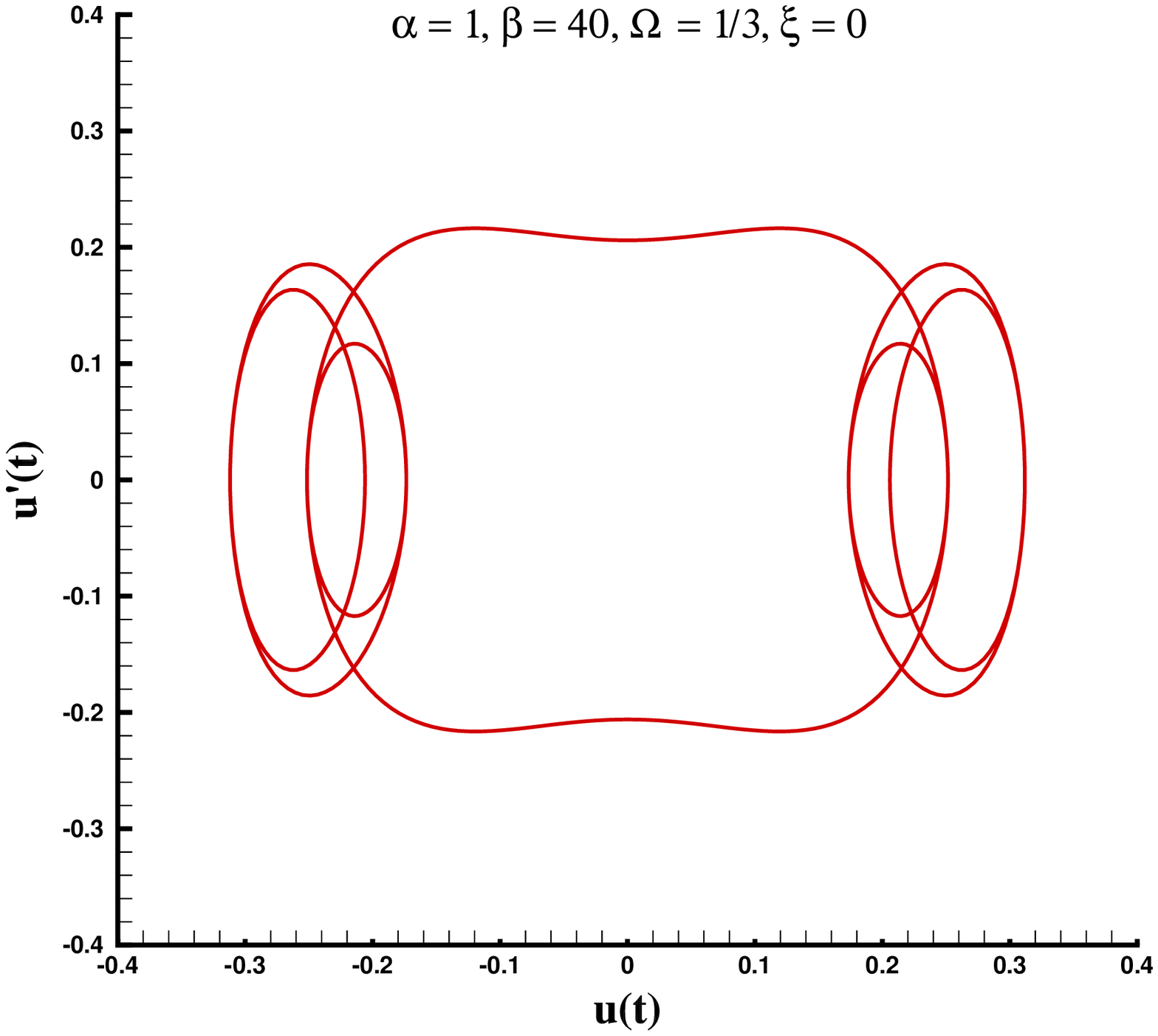}}
             \subfigure[]{\includegraphics[width=2.5in]{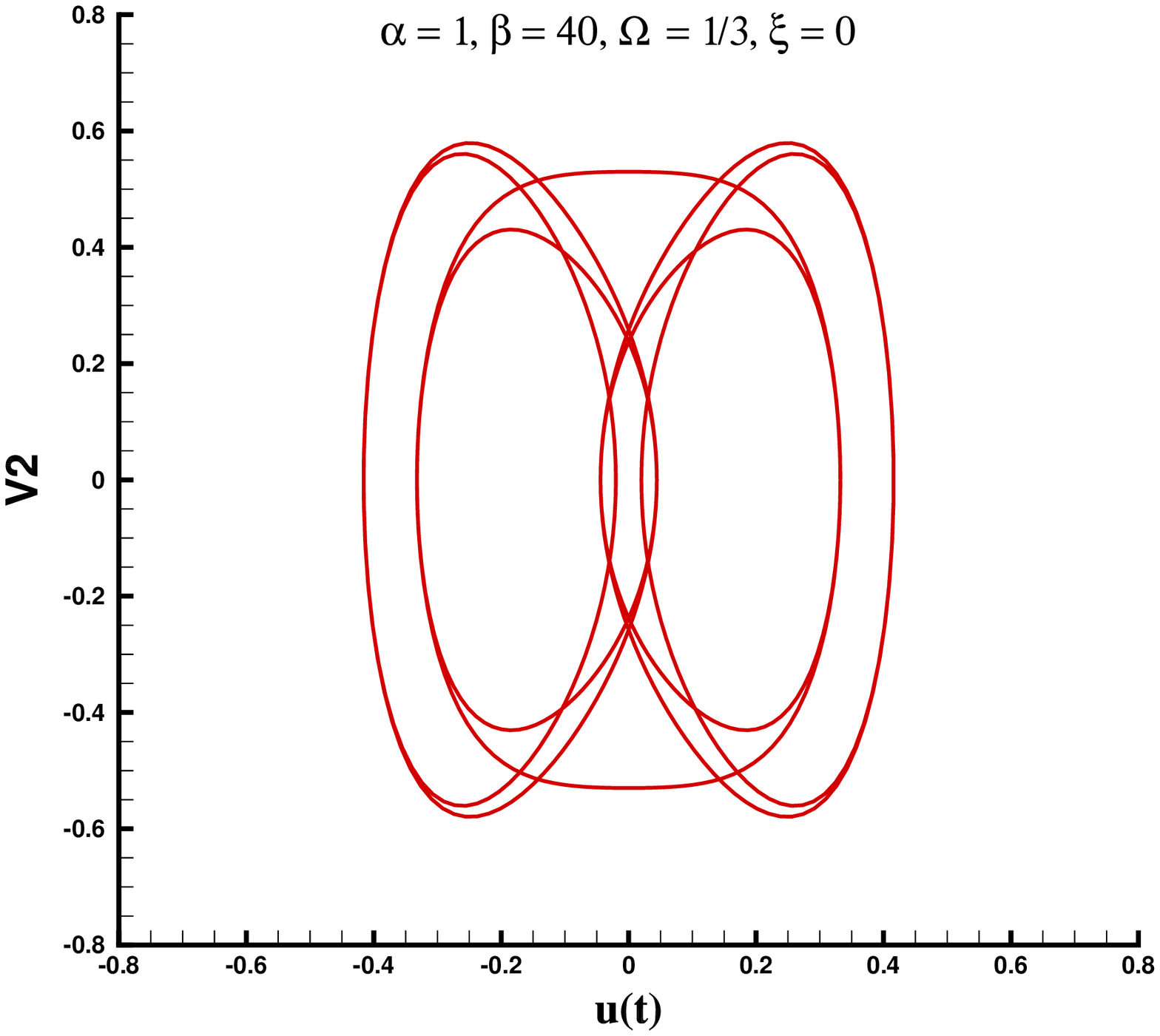}}
        \end{tabular}
 \end{center}
 \caption{Multiple limit-cycles of the Duffing equation (\ref{geq:original}) in cases of $\alpha=1,\Omega=1/3, \xi = 0$ and $\beta=40$, given by the HAM iteration approach described in \S~4.2 when $\lambda=\omega_{1}$ and $\delta=\omega_{4}^{2}-\omega_{1}^{2}$.  (a)  using  the initial guess (\ref{u[0]:beta=40:xi=0:K=4:a}) and $c_{0}=-3/2$; (b) using the initial guess (\ref{u[0]:beta=40:xi=0:K=4:b}) and   $c_{0}=-4/5$; (c)  using  the initial guess (\ref{u[0]:beta=40:xi=0:K=4:c}) and $c_{0}=-4/5$; (d) using the initial guess (\ref{u[0]:beta=40:xi=0:K=4:d}) and $c_{0}=-1$.  }
 \label{fig:limit-cycle:beta40:K4}
\end{figure}

In this case, we have $\lambda=\omega_{1}$ and $\delta = \omega_{4}^{2}-\omega_{1}^{2}$  so that the kernel of the corresponding auxiliary linear operator $\cal L$ is a vector space of {\bf 8} dimensions, say,
\begin{equation}
\ker[{\cal L}] = \sum_{n=1}^{4} \left[  A_{n} \cos(\omega_{n} t) +  B_{n} \sin(\omega_{n} t)  \right],  \label{ker[L]:kappa=4}
\end{equation}
where $A_{n} $ and $B_{n} $ are arbitrary constants.

 According to (\ref{u[0]:B}), the initial guess should be in the form
 \begin{equation}
 u_{0}(t) = \sum_{n=1}^{4} \Big[ a_{0,n} \cos(\omega_{n} t) + b_{0,n} \sin(\omega_{n} t) \Big]. \label{u[0]:B:K=4}
 \end{equation}
 Substituting it into the forced Duffing equation (\ref{geq:original}) and enforcing the coefficients of the terms $\cos(\omega_{n}t)$ and $\sin(\omega_{n}t)$ $(n=1,2,3,4)$ being zero, we have a set of eight nonlinear algebraic equations, whose real solutions determine the eight unknown constants in (\ref{u[0]:B:K=4}).  It is found that the set of these eight nonlinear algebraic equations has {\bf four} real solutions in the case of  $\alpha=1,  \Omega=1/3$, $\xi=0$ and $\beta=5$:
 \begin{eqnarray}
&& a_{0,1}=0.261766, a_{0,2}=0.0766644, a_{0,3} = -0.563565, a_{0,4}=-0.0899237; \hspace{1.5cm}\label{u[0]:beta=5:xi=0:K=4:a} \\
&&  a_{0,1}=0.524251, a_{0,2}=-0.119311, a_{0,3} = 0.150425, a_{0,4}= 0.0487805; \label{u[0]:beta=5:xi=0:K=4:b} \\
&&   a_{0,1}=0.420975, a_{0,2}=0.0508348, a_{0,3} = -0.278947, a_{0,4}=-0.0849981; \label{u[0]:beta=5:xi=0:K=4:c} \\
&&   a_{0,1}=0.104111, a_{0,2}=0.000412556, a_{0,3} = -0.00628524, a_{0,4}=1.07865, \label{u[0]:beta=5:xi=0:K=4:d}
 \end{eqnarray}
where $b_{0,n}=0$ for $n=1,2,3,4$, corresponding to the four initial guesses in the form (\ref{u[0]:B:K=4}).

Using the initial guess (\ref{u[0]:beta=5:xi=0:K=4:a}) and the convergence-control parameter $c_{0}=-2/3$, we gain a convergent  series  solution of the  limiting cycle, which is exactly the same as  (a) in Figure~\ref{fig:limit-cycle:K3}, by means of the 2nd-order HAM iteration: the residual error square of the forced Duffing equation (\ref{geq:original}) decreases from  0.14 at the beginning to  $ 3.6  \times 10^{-30}$ at the 30th iteration.

Using the initial guess (\ref{u[0]:beta=5:xi=0:K=4:b}) and the convergence-control parameter $c_{0}=-1$, we gain a convergent  series  solution of the  limiting cycle, which is exactly the same  as (b) in Figure~\ref{fig:limit-cycle:K3}, by means of the 2nd-order HAM iteration: the residual error square of the forced Duffing equation (\ref{geq:original}) decreases from  0.021 at the beginning to  $ 8.4  \times 10^{-30}$ at the 15th iteration.

Using the initial guess (\ref{u[0]:beta=5:xi=0:K=4:c}) and $c_{0}=-3/2$, we gain a convergent  series  solution of the  limiting cycle, which is exactly the same as (c) in Figure~\ref{fig:limit-cycle:K3},  by means of the 2nd-order HAM iteration: the residual error square of the forced Duffing equation (\ref{geq:original}) decreases from  0.022 at the beginning to  $6.6 \times 10^{-30}$ at the 20th iteration.

Using the initial guess (\ref{u[0]:beta=5:xi=0:K=4:d}) and $c_{0}=-1$, we gain a {\em new} convergent  series  solution, shown as (d) in Figure~\ref{fig:limit-cycle:K3},   by means of the 2nd-order HAM iteration: the residual error square of the forced Duffing equation (\ref{geq:original}) decreases from  1.48  at the beginning to  $5.6 \times 10^{-16}$ at the 30th iteration.  It is interesting that this  is a {\em new} solution,  which is however not found by means of $\lambda=\omega_{1}$ and  $\kappa = 3$.  This is mainly because, when $\kappa = 4$,  we should solve two more nonlinear algebraic equations to gain the initial guess than the case of $\kappa=3$.  This leads  to  one more  initial  guess that  gives  one more  limit-cycle by means of the HAM approach described in this paper.     It seems that, in the frame of the HAM described in \S~2 and \S~3,  the larger the value of $\delta$, the greater the possibility to find multiple solutions (if they indeed exist).   This  further  shows   the  validity and novelty  of  our  HAM approach beyond perturbation.

As shown in Figure~\ref{fig:limit-cycle:K3:high-nonlinearity}, for a given $\beta\in[10,40]$ (corresponding to high nonlinearity),  only one   limit-cycle is found by the HAM approach using $\lambda=\omega_{1}$ and $\delta = \omega_{3}^{2}-\omega_{1}^{2}$,  corresponding to $\kappa = 3$.  Do multiple limit-cycles exist  in high nonlinearity, say, for a large $\beta$?   Without loss of generality, let us consider
the case of $\alpha=1, \Omega=1/3,\xi=0$ and $\beta=40$.  It is found that, when $\kappa=4$,   we have the  {\em four} corresponding initial guesses in the form of (\ref{u[0]:B:K=4}):
\begin{eqnarray}
&& a_{0,1} = 0.131123,a_{0,2} = -0.00645067,a_{0,3} = -0.0312269,a_{0,4} = 0.336815; \hspace{1.5cm}\label{u[0]:beta=40:xi=0:K=4:a}\\
&& a_{0,1} = 0.287691,a_{0,2} = -0.0231836,a_{0,3} = -0.0268927,a_{0,4} = 0.0697927; \label{u[0]:beta=40:xi=0:K=4:b} \\
&& a_{0,1} = 0.296537,a_{0,2} = -0.0580039,a_{0,3} = 0.0566195,a_{0,4} = -0.0810271; \label{u[0]:beta=40:xi=0:K=4:c} \\
&& a_{0,1} = 0.261690,a_{0,2} = -0.0276672,a_{0,3} = 0.0714539,a_{0,4} = -0.156464; \label{u[0]:beta=40:xi=0:K=4:d}
\end{eqnarray}
where $b_{0,n} = 0$ $(n= 1,2,3,4)$.

  Using the initial guess (\ref{u[0]:beta=40:xi=0:K=4:a}) and $c_{0} =-3/2 $, we gain a convergent series solution, shown as (a) in Figure~\ref{fig:limit-cycle:beta40:K4},  by means of the 5th-order HAM iteration: the residual error square of the forced Duffing equation ({\ref{geq:original}) decreases from  0.26 at the beginning to $10^{-21}$  at the 10th iterations.

  Using the initial guess (\ref{u[0]:beta=40:xi=0:K=4:b}) and $c_{0} =-4/5 $, we gain a convergent series solution, shown as (b) in Figure~\ref{fig:limit-cycle:beta40:K4}, by means of the 5th-order HAM iteration: the residual error square of the forced Duffing equation ({\ref{geq:original}) decreases from  0.016 at the beginning to $2.7 \times 10^{-17}$  at the 10th iterations. This limit-cycle is  exactly  the same as  (d)  in  Figure~\ref{fig:limit-cycle:K3:high-nonlinearity}, the only one limit-cycle  when $\beta=40$  given   by means of  $\lambda=\omega_{1}$ and $\delta=\omega_{3}^{2}-\omega_{1}^{2}$, corresponding to $\kappa=3$.

  Using the initial guess (\ref{u[0]:beta=40:xi=0:K=4:c}) and $c_{0} =-4/5 $, we gain a convergent series solution, shown as (c) in Figure~\ref{fig:limit-cycle:beta40:K4},  by means of the 5th-order HAM iteration: the residual error square of the forced Duffing equation ({\ref{geq:original}) decreases from  0.013 at the beginning to $8.1 \times 10^{-17}$  at the 10th iterations.

 Using the initial guess (\ref{u[0]:beta=40:xi=0:K=4:d}) and $c_{0} =-1 $, we gain a convergent series solution, shown as (d) in Figure~\ref{fig:limit-cycle:beta40:K4},  by means of the 5th-order HAM iteration: the residual error square of the forced Duffing equation ({\ref{geq:original}) decreases from  0.044 at the beginning to $ 6.4  \times 10^{-21}$  at the 10th iterations.

Note that when $\kappa=3$, i.e. $\lambda=\omega_{1}$ and $\delta = \omega_{3}^{2}-\omega_{1}^{2}$,  we found only one limit-cycle for $\beta\in[10,40]$.  However, when $\kappa=4$, i.e. $\lambda=\omega_{1}$ and $\delta = \omega_{4}^{2}-\omega_{1}^{2}$, we successfully  gain four limit-cycles in the case of $\beta=40$, corresponding to a very high nonlinearity.
 It seems that, the larger the value of $\kappa$, say, the larger of $\delta$,  more limit-cycles of the forced Duffing equation (\ref{geq:original}) could be found.  Note that $\beta=40$ corresponds to a high nonlinearity: this verifies the validity of our HAM approach for high nonlinearity.  This is one of advantages of the HAM, which has been proved in many articles (for example, please refer to Zhong and Liao \cite{Zhong2018JFM}).
  In summary,  all of these results illustrate the validity and novelty of our HAM approach described in \S~2 and \S~3.  Note that  the so-called ``small denominator problem'' never appears for the forced Duffing equation (\ref{geq:original}) by means of the HAM approach.

 Note that, according to traditional mathematical theories, a linear differential operator $\cal L$, whose kernel
is the same as the vector space of {\bf 8} dimension defined by (\ref{ker[L]:kappa=4}), should correspond to the 8th-order differential equation
\begin{eqnarray}
{\cal L}[u] &=&u^{(8)} + \left( \omega_{1}^{2}+\omega_{2}^{2}+\omega_{3}^{2}+\omega_{4}^{2}\right)u^{(6)} \nonumber\\
&+& \left(  \omega_{1}^{2} \omega_{2}^{2}+\omega_{1}^{2} \omega_{3}^{2} +\omega_{1}^{2} \omega_{4}^{2}+\omega_{2}^{2} \omega_{3}^{2}++\omega_{2}^{2} \omega_{4}^{2}++\omega_{3}^{2} \omega_{4}^{2}\right) u^{(4)} \nonumber\\
&+& \left(  \omega_{1}^{2} \omega_{2}^{2} \omega_{3}^{2} +\omega_{1}^{2} \omega_{2}^{2} \omega_{4}^{2} + \omega_{1}^{2} \omega_{3}^{2} \omega_{4}^{2}+\omega_{2}^{2} \omega_{3}^{2} \omega_{4}^{2}\right)u'' \nonumber\\
&+& \omega_{1}^{2} \omega_{2}^{2} \omega_{3}^{2} \omega_{4}^{2} u= 0, \label{L:kappa=4}
\end{eqnarray}
whose inverse operator reads
 \begin{eqnarray}
&&{\cal L}^{-1}\Big[ A \cos(\omega t) + B \sin(\omega t) \Big] \nonumber\\
&=& \frac{A \cos(\omega t) + B \sin(\omega t)}{(\omega^{2}-\omega_{1}^{2})(\omega^{2}-\omega_{2}^{2})(\omega^{2}-\omega_{3}^{2})(\omega^{2}-\omega_{4}^{2})},  \hspace{0.5cm} \omega \notin\left\{ \omega_{1}, \omega_{2}, \omega_{3}, \omega_{4}\right\} \hspace{1.0cm} \label{L:inverse:kappa=4}
\end{eqnarray}
for arbitrary constants $A$ and $B$. However,  the above expression is obviously {\em different} from our inverse operator (\ref{L:inverse:2nd:C}) that  looks like one for a 2nd-order linear differential equation!  In fact, we even do {\em not} know how to explicitly express the corresponding auxiliary linear operator $\cal L$ when $\kappa=4$, say,  $\lambda=\omega_{1}$ and $\delta=\omega_{4}^{2}-\omega_{1}^{2}$, but fortunately it is {\em unnecessary} to know it in the frame of the HAM.   The most important fact  is  that our HAM-based approach is valid and the corresponding solution series of the  limiting cycles converge quickly, as mentioned above, which verifies the validity and novelty of our HAM approach mentioned in \S~2 and \S~3.

 \section{Discussions and concluding remarks}

First of all, when perturbation method is used to solve the forced Duffing equation (\ref{geq:original}), the so-called ``small  denominator  problem'' is {\em unavoidable} when $\omega \to 1$ and $\xi\to 0$, which leads to the divergence of the perturbative series even for rather small $\beta$, corresponding to  a very weak nonlinearity.  However,  for the HAM approach described in \S~2 and \S3,  such kind of  small  denominators {\bf never} appear for arbitrary values of  physical parameters $\alpha, \beta, \omega$ and $\xi$ so  that  the  so-called  ``small  denominator  problem'' {\bf never} occurs!    Note that even in the case of large $\beta$, corresponding to high nonlinearity, multiple  limit-cycles are successfully found by our HAM approach.  All of these illustrate the validity and novelty of the HAM  approach.  Thus, from the viewpoint of the HAM approach described in this paper,  the so-called  ``small  denominator  problem''  does {\bf not} really exist!  This  suggests that whether  or not the so-called  ``small  denominator  problem''  really exists should highly depend on the used method:  it indeed exists for perturbation methods, but {\bf not} for the HAM!  Thus,   the origin of the so-called   ``small  denominator  problem''  comes from the limitations and restrictions of perturbation method as a methodology.      In other words,  the ``small denominator problem''  is only an {\em artifact} of perturbation method.  Thus, abandoning perturbation method but using the HAM,  we can completely avoid the ``small  denominator  problem''.    Note that the ``small  denominator  problem'' has been regarded as a huge obstacle for many  open  problems in science.  So, the HAM provides us a new way to attack them.

Secondly,  unlike all other approximation techniques (including perturbation methods), we can directly {\bf define} the inverse operator ${\cal L}^{-1}$  of an undetermined linear operator in the frame of the HAM  so as to easily gain the solutions of the linear high-order equations.  It should be emphasized that it is the HAM that provides us such kind of great freedom \cite{Liao2007SAM,Liao2016NA}.  Using such kind of freedom, the so-called ``small  denominator  problem''  can be completely avoided, as illustrated in this paper.
Note that, according to traditional mathematical theories, a linear differential operator $\cal L$, whose kernel
is a vector space of {\bf 4} dimension defined by (\ref{ker[L]:kappa=2}),  should correspond to the 4th-order linear differential equation (\ref{L:kappa=2}),  whose inverse operator should be expressed by (\ref{L:inverse:kappa=2}).
Similarly, a linear differential operator $\cal L$, whose kernel is a vector space of {\bf 6} dimension defined by (\ref{ker[L]:kappa=3}),  should correspond to the 6th-order  linear differential equation (\ref{L:kappa=3}),  whose inverse operator should be expressed by (\ref{L:inverse:kappa=3}).
In addition, a linear differential operator $\cal L$, whose kernel is a vector space of {\bf 8} dimension defined by (\ref{ker[L]:kappa=4}),  should correspond to the 8th-order  linear differential equation (\ref{L:kappa=4}),  whose inverse operator should be expressed by (\ref{L:inverse:kappa=4}). However, when $\lambda=\omega_{1}$ and $\delta = \omega^{2}_{\kappa} -\omega^{2}_{1} $,  although its kernel defined by (\ref{L:kernel})  is a vector space of {\bf 4, 6, 8} dimension for $\kappa = 2,3,4$, respectively, its inverse operator defined by (\ref{L:inverse}) always looks like that of a second-order linear operator whose kernel should be a vector space of {\bf 2} dimension according to the traditional mathematical theorms!   Obviously,  the inverse operator (\ref{L:inverse}), which  we directly {\em define} in the frame of the HAM, is quite {\em different}  from (\ref{L:inverse:kappa=2}), (\ref{L:inverse:kappa=3}) and (\ref{L:inverse:kappa=4}).  Not that we even do {\em not} know how to explicitly express its corresponding auxiliary linear operator $\cal L$.  Fortunately, it is {\em unnecessary}  to know the undetermined linear operator $\cal L$ in the frame of the HAM.  Thus,  to the best of author's knowledge,  the auxiliary linear operator defined by (\ref{L:inverse}) and (\ref{L:kernel}) is fundamentally different from all known traditional linear operators.  Note that,  in the previous applications of the HAM \cite{Liao1992PhD, Liao2003Book, Liao2012Book, Liao1995IJNLM, Liao1997IJNLM, Liao1999IJNLM, Liao2004AMC, Liao2007SAM,  Liao2009CNSNS, Liao2010CNSNS, Liao2016JFM, Liao2020SCPMA, Zhong2018JFM, Zhu2006QF,  Bouremel2007CNSNS,  Nassar2011,  Kimiaeifar2011CMA,  Ghotbi2011,  Duarte2015CSF,  Sardanyes2015,  VanGorder2017, Pfeffer2017,  Cullen2019JCP,  Sultana2019EPJP,  Massa2020,  Kaur2022JMAA,  Masjedi2022AMM,  Botton2022AMM}, one mostly chooses a proper linear auxiliary operator $\cal L$ and then find its  corresponding inverse operator ${\cal L}^{-1}$ so as to solve the high-order equations.  However, in this paper, we directly {\em define}  the inverse operator  ${\cal L}^{-1}$  but  do {\em not} care about the explicit expression of the corresponding auxiliary linear operator $\cal L$ at all.  This might be a breakthrough in the field of differential equations.  It further illustrates the novelty and great potential of the so-called ``method of directly defining inverse mapping'' (MDDiM), which was proposed  by Liao and Zhao \cite{Liao2016NA} in the frame of the HAM and  has been successfully applied to solve many types of nonlinear equations   \cite{ KV2018NA, KV2018AMC, KV2019CNA,  thesis2019Dewasurendra,  KV2021AMNS, KV2021CAM, KV2022CMA, Nave2018JBD, Nave2018JMM, Nave2019JMM, Nave2020BS,  Munjam2022IJAE, Munjam2023IJAE}.

Thirdly,  unlike perturbation techniques, the HAM provides us great freedom to choose initial guesses.  Using such kind of freedom,  we can gain the multiple limit-cycles of the  forced  Duffing equation (\ref{geq:original}) by means of the HAM.  Note that,  when $\lambda=\omega_{1}$ and $\delta = \omega^{2}_{\kappa} -\omega^{2}_{1} $ for $\kappa\geq 2$,  the larger the value of $\delta$ in the definitions (\ref{L:inverse}) and (\ref{L:kernel}), the greater the probability to find more limit-cycles of the forced Duffing equation (\ref{geq:original}).  In contrast,  perturbation method provides only one initial guess (\ref{initial:pert}) and thus at most one limit-cycle.  Thus, this illustrates  the novelty of our HAM approach and its advantages beyond perturbation.

Unlike all other approximation techniques (including perturbation methods), the HAM contains the so-called convergence-control parameter $c_{0}$, which  provides a simple way to guarantee the convergence of solution series even when the nonlinearity is very high, as illustrated in this paper and also in other publications about the HAM \cite{Zhu2006QF,  Bouremel2007CNSNS,  Nassar2011,  Kimiaeifar2011CMA,  Ghotbi2011,  Duarte2015CSF,  Sardanyes2015,  VanGorder2017, Pfeffer2017,  Cullen2019JCP,  Sultana2019EPJP,  Massa2020,  Kaur2022JMAA,  Masjedi2022AMM,  Botton2022AMM}.   This  guarantees that the HAM-based approach is generally valid for high nonlinearity.

As pointed out by Giorgilli \cite{Giorgilli1998}, Duffing equation ``is perhaps the simplest example of a non-integrable system exhibiting {\bf all} problems due to the small denominators''.   So, although the forced Duffing equation (\ref{geq:original}) is used here as an example to illustrate the validity and novelty of the HAM approach and its advantages beyond perturbation, most conclusions mentioned above have general meanings.

What will happen if  the homotopy analysis method (HAM) instead of perturbation method is first proposed by an intelligent being on a planet in the universe?   Certainly, using the HAM-based approach mentioned in this paper, this kind of intelligent being should have {\em no} ideas of  ``small denominator  problem'' at all!
Thus, the famous ``small  denominator  problem'' does {\em not} really exist and should be an {\em artifact} of perturbation method.  Therefore, completely abandoning perturbation methods but using the HAM-based MDDiM,  we can thoroughly  avoid  ``small  denominator  problems''  and besides could  attack many open problems related to  small  denominators.   In addition,  we illustrated here that a nonlinear differential equation can be solved by directly defining a proper inverse operator of an undetermined linear operator.  Hopefully,  this fact might lead to a breakthrough in the field of differential equations.

In summary, completely  abandoning  perturbation methods but using the HAM-based MDDiM, one would be never troubled by  ``small denominator problems''!

\section*{Acknowledgements}  This work is partly supported by National Natural Science Foundation of China (Approval No. 12272230) and Shanghai Pilot Program for Basic Research - Shanghai Jiao Tong University ( No. 21TQ1400202).

\section*{Data Availability Statements}
The author confirms that the data supporting the findings of this study are available within the article.

\section*{Author contributions}
S.L. is the unique author of this article.

\section*{Declaration of competing interest} The author declares that he has no competing financial interests.



\end{document}